%% file: sample-acmsmall.tex
\documentclass[acmsmall]{acmart}
\AtBeginDocument{%
  \providecommand\BibTeX{{%
    \normalfont B\kern-0.5em{\scshape i\kern-0.25em b}\kern-0.8em\TeX}}}

\setcopyright{acmlicensed}
\copyrightyear{2018}
\acmYear{2018}
\acmDOI{XXXXXXX.XXXXXXX}

\acmJournal{JACM}
\acmVolume{37}
\acmNumber{4}
\acmArticle{111}
\acmMonth{8}

\usepackage{algorithm}
\usepackage{algorithmic}
\usepackage{bibentry}
\usepackage{amsfonts}
\usepackage{subfigure}
\usepackage{multirow}
\usepackage{amsmath}
\usepackage{graphicx}
\usepackage{epsfig}
\usepackage{color}
\usepackage{epstopdf}
\usepackage{rotating}
\usepackage{caption}
\usepackage{float}
\usepackage{multicol}
\usepackage{graphicx}
\usepackage{bbding}
\usepackage{algorithm}
\usepackage{algorithmic}
\usepackage{balance}
\usepackage{microtype}
\usepackage{booktabs}
\usepackage{amsmath}
\usepackage{amsmath, bm}

\usepackage{mathtools}
\usepackage{etoolbox}
\usepackage{cases}
\usepackage{enumitem}
\usepackage{xcolor}
\usepackage{xcolor}
\usepackage{color}

\usepackage{cases}

\usepackage{amsmath,amsfonts}
\usepackage{algorithmic}
\usepackage{graphicx}
\usepackage{textcomp}

\usepackage{mathtools}

\usepackage{amsthm}
\usepackage{color}
\usepackage{multirow}

\usepackage{cases}
\usepackage{amsmath,amsfonts}
\usepackage{algorithmic}
\usepackage{graphicx}
\usepackage{textcomp}
\usepackage{xcolor}

\definecolor{brown}{RGB}{139,64,0}
\definecolor{pink}{RGB}{255,170,182}
\definecolor{purple}{RGB}{160,32,240}

\usepackage{cases}
 
\usepackage{amsmath,amsfonts}
\usepackage{algorithmic}
\usepackage{graphicx}
\usepackage{textcomp}
\usepackage{xcolor}

\begin{document}

\title{Multi-agent Attacks for Black-box Social Recommendations}


\author{Shijie Wang}
\affiliation{%
  \institution{The Hong Kong Polytechnic University}
  \country{Hong Kong}}
\email{shijie.wang@connect.polyu.hk}

\author{Wenqi Fan}
\affiliation{%
  \institution{The Hong Kong Polytechnic University}
  \country{Hong Kong}}
\email{wenqifan03@gmail.com}

\author{Xiao-yong Wei}
\affiliation{%
  \institution{The Hong Kong Polytechnic University}
  \country{Hong Kong}}
\email{x1wei@polyu.edu.hk}

\author{Xiaowei Mei}
\affiliation{%
  \institution{The Hong Kong Polytechnic University}
  \country{Hong Kong}}
\email{michael.mei@polyu.edu.hk}

\author{Shanru Lin}
\authornote{Corresponding author: Shanru Lin, City University of Hong Kong, Hong Kong.}
\affiliation{%
  \institution{City University of Hong Kong}
  \country{Hong Kong}}
\email{lllam32316@gmail.com}

\author{Qing Li}
\affiliation{%
  \institution{The Hong Kong Polytechnic University}
  \country{Hong Kong}}
\email{csqli@comp.polyu.edu.hk}

\input{sections/abs}



\maketitle

\input{sections/introduction-V2}

\input{sections/preliminary-V2}

\input{sections/model-V2}

\input{sections/model-V2-part2}
\input{sections/experiments-V2}
\input{sections/relatedwork}

\input{sections/conclusion}

\bibliographystyle{ACM-Reference-Format}
\bibliography{references/references}

\end{document}

%% file: sections/abs.tex
\begin{abstract}

The rise of online social networks has facilitated the evolution of social recommender systems, which incorporate social relations to enhance users' decision-making process. With the great success of Graph Neural Networks (GNNs) in learning node representations, GNN-based social recommendations have been widely studied to model user-item interactions and user-user social relations simultaneously. Despite their great successes, recent studies have shown that these advanced recommender systems are highly vulnerable to adversarial attacks, in which attackers can inject well-designed fake user profiles to disrupt recommendation performances.  While most existing studies mainly focus on \emph{targeted attacks} to promote target items on vanilla recommender systems,  \emph{untargeted attacks} to degrade the overall prediction performance are less explored on social recommendations under a \emph{black-box} scenario. To perform untargeted attacks on social recommender systems, attackers can construct malicious social relationships for fake users to enhance the attack performance. However, the coordination of social relations and item profiles is challenging for attacking black-box social recommendations. To address this limitation, we first conduct several preliminary studies to demonstrate the effectiveness of cross-community connections and cold-start items in degrading recommendations performance.  Specifically, we propose a novel framework \emph{MultiAttack} based on multi-agent reinforcement learning to coordinate the generation of cold-start item profiles and cross-community social relations for conducting untargeted attacks on black-box social recommendations.  Comprehensive experiments on various real-world datasets demonstrate the effectiveness of our proposed attacking framework under the black-box setting.

\end{abstract}

%% file: sections/introduction-V2.tex
 \section{Introduction}
In the past few decades, recommender systems as intelligent filtering tools have played an increasingly important role in people’s daily lives~\cite{tang2016recommendation, fan2022comprehensive, cheng2016wide}, such as online shopping (e.g., Taobao, Amazon), social media (e.g., Twitter, Facebook), job matching (e.g., LinkedIn), etc. 
To further enhance recommendation performance,  the use of social relations has received increasing attention due to their significant influences on users' decisions~\cite{ma2008sorec, ma2009learning}, which is known as \emph{social recommendations}~\cite{fan2019graph, tang2016recommendation}. 
In recent years, by modeling complex user-item interactions and social relations ~\cite{fan2018deep,wu2019neural}, recommender systems have been further improved with the rapid advancement and widespread applications of deep learning techniques~\cite{fan2019deep_dscf, fan2019graph,qu2024tokenrec}, specifically Graph Neural Networks (GNNs) and Recurrent Neural Networks (RNNs).
While these techniques have brought powerful representation learning capabilities in recommender systems ~\cite{chen2022knowledge, song2020poisonrec,zhang2024graph}, recent studies have found that deep learning-based recommender systems are highly susceptible to adversarial attacks by generating fake users with well-designed profiles (i.e., a set of items) ~\cite{fan2021attacking, song2020poisonrec}.

The majority of existing attacking methods focus on white-box or grey-box attacks~\cite{fang2020influence,christakopoulou2019adversarial}, which require complete or partial user interaction history and recommendation model knowledge ~\cite{li2016data, tang2020revisiting}. 
Considering the protection of privacy and security, 
adversarial black-box attacks, which aim to generate well-designed user profiles by querying the black-box systems for rewards~\cite{chen2022knowledge, song2020poisonrec, fan2021attacking}, are more practical.
To be specific, existing advanced approaches, such as PoisonRec ~\cite{song2020poisonrec} and KGAttack ~\cite{chen2022knowledge},  generate fake users' item profiles by querying target recommender systems. 
CopyAttack leverages cross-domain user behaviors to copy `real' user profiles to attack black-box systems~\cite{fan2021attacking,fan2023adversarial}.
Although some studies have successfully attacked black-box recommender systems, they are limited in social recommender systems as they ignore the impact of social relations. 
Meanwhile, most existing attacking methods primarily address \emph{targeted attacks} with the goal to promote or demote certain items (i.e., target items), while \textbf{untargeted attacks} that aim to degrade the overall recommendation performance are still less explored. 
However, it is worth noting that both untargeted and targeted attacks are crucial for investigating the vulnerability of social recommender systems. Untargeted attacks can lead to poor users' online experiences\&trust in the systems, and even severe economic losses, by recommending large amounts of irrelevant and low-quality items to users. For example, suppose there are two competitive online shopping platforms, and one platform launches an attack against the other. In such a scenario, it is more effective to reduce the experience and satisfaction of all users, rather than promoting specific products solely. 
Additionally, to build more robust recommender systems, they typically need to conduct adversarial training, which also requires attacks first to minimize the recommendation performance.
Therefore, it is necessary to study untargeted black-box attacks in social recommendations.

\begin{figure}[tb]
 \centering
\subfigure{\label{fig:rl}\includegraphics[width=0.56\linewidth]{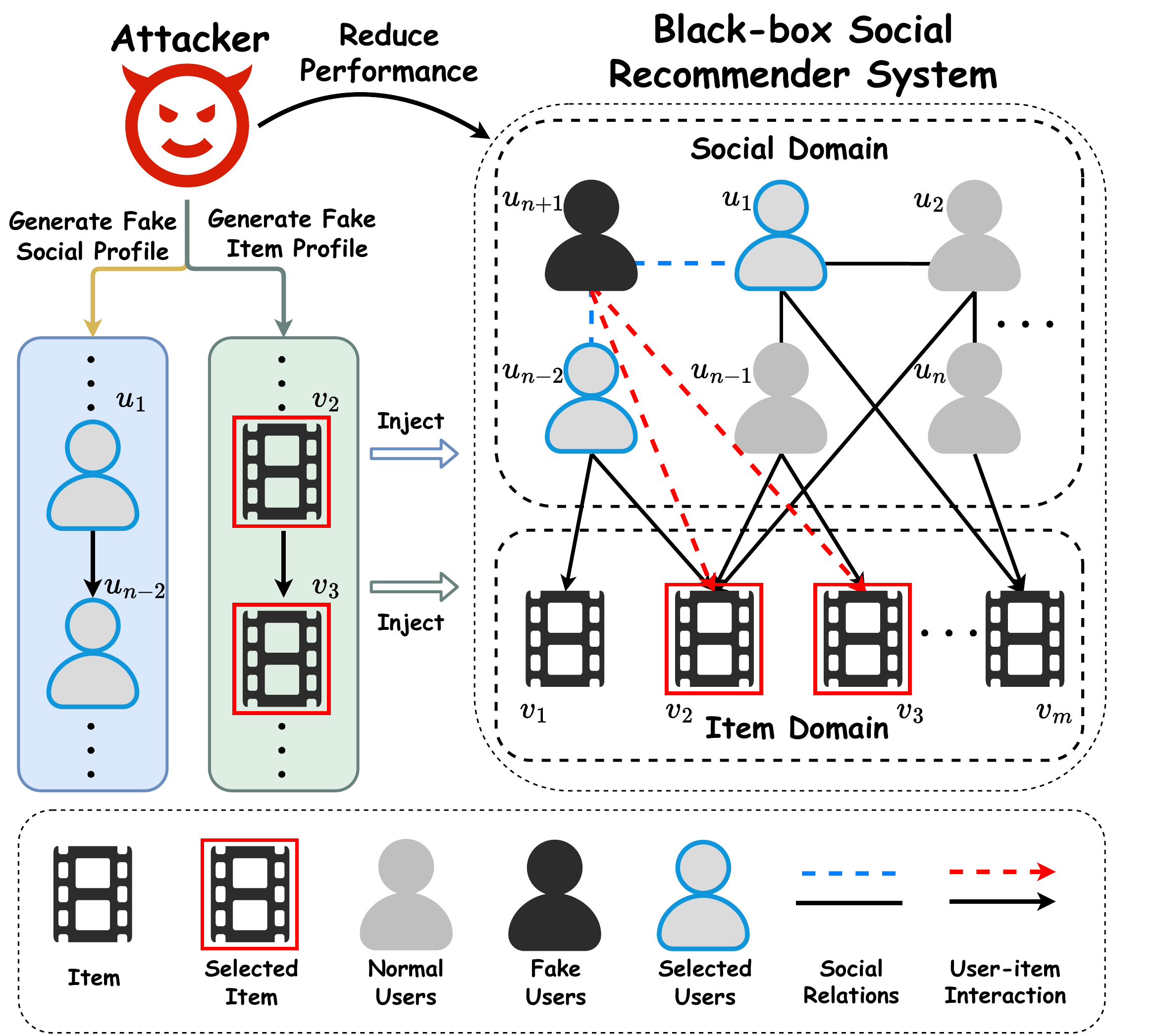}}
\caption{Illustration of untargeted black-box attacks in social recommender system. The objective of attackers is to reduce the overall recommendation performance by injecting a fake user (i.e., $u_{n+1}$) with profiles,  consisting of \emph{fake item profile} (i.e., $\{..., v_2, v_3, ...\}$) and \emph{fake social profile} (i.e., $\{..., u_1, u_{n-2}, ...\}$).
}
\label{fig:Illustration}
\vskip -0.218in
\end{figure}

Attacking social recommender systems under the black-box setting presents significant and multifaceted challenges. 
In order to reduce the overall performance of social recommender systems, the objective of the attacker is to generate \emph{item profiles} and \emph{social profiles} simultaneously, as shown in Figure \ref{fig:Illustration}. 
Different from targeted attacks with certain targeted items, evaluating the overall recommendation performance of target systems under the black-box setting is less explored.  An appropriate strategy on evaluating untargeted attacks can guide the generation of item profiles from a large-scale item set. 
Hence, the first challenge is how to effectively generate high-quality item profiles for untargeted black-box attacks, so as to reduce the overall recommendation performance.  
Moreover, since users' social relations can significantly impact their decision making in social recommendations~\cite{fan2019graph}, it is imperative for attackers to take advantage of users'  social connections among users by establishing influential \emph{social profiles}. 
Thus, the second challenge is how to effectively generate high-quality social profiles in attacking social recommendations.
Furthermore, the generation of item profiles and social profiles poses a complex combinatorial optimization problem due to their intricate interlinking. 
Optimizing item and social profiles independently could easily lead to sub-optimal attacking performance on social recommendations. 
Therefore, the third challenge is how to coordinate the generation of social and item profiles for attacking black-box social recommendations.

To address the aforementioned challenges, in this paper, we propose a novel attacking framework \textbf{MultiAttack} for black-box social recommender systems. 
In particular, we first conduct comprehensive preliminary experiments to investigate the strategies (i.e., cold-start items and cross-community users) to generate item and social profiles for untargeted black-box attacks on social recommendations. 
To achieve coordination of the generation of item profiles and social profiles, we propose a multi-agent reinforcement learning attacking framework by developing multiple agents guided by a centralized critic network to generate cross-community social profiles and cold-start item profiles, so as to degrade the overall performance of the black-box social recommendations. 
The main contributions of this paper can be summarized as follows:
\begin{itemize}[leftmargin=*]

\item To the best of our knowledge, this is the first work to investigate untargeted black-box attacks for social recommendations with the goal of degrading the overall recommendation performance.

\item We introduce a principle strategy to generate fake user profiles for attacking social recommendations, where cold-start items and cross-community users are leveraged to generate high-quality item and social profiles, respectively.

\item We propose a novel attacking framework (\textbf{MultiAttack}) for attacking the black-box social recommendation, where multiple agents are introduced to coordinate the generation of social profiles and item profiles based on multi-agent reinforcement learning.

\item  Comprehensive experiments on various real-world datasets are conducted to demonstrate the effectiveness of the proposed attacking framework.

\end{itemize}

%% file: sections/preliminary-V2.tex
\section{Preliminary Study}

In order to address the challenges in degrading the overall performance of social recommendations, in this section, we carry out a set of preliminary studies on LastFM and Ciao (see Section \ref{datasets} for the details of the datasets).
More specifically, 
we investigate the strategies to generate item and social profiles for untargeted black-box attacks.

\subsection{Cold-start Items for Untargeted Black-box Attacks} \label{cold-start}
Compared with most existing targeted attacks which are easily executed on a set of target items for generating fake item profiles, 
it is highly challenging to conduct untargeted black-box attacks to effectively generate high-quality fake item profiles from a large-scale item set.
It is worth mentioning that most users' online behaviours in recommender systems are following the power law  distribution~\cite{celma2008new,abdollahpouri2017controlling}, also known as the 80/20 rule (i.e., Pareto Principle), in which a small proportion (e.g., less than 20\%) of items (i.e., popular items) often account for a large proportion of users' online behaviours (e.g., more than 80\%)~\cite{brynjolfsson2011goodbye}, while cold-start items have a few interactions from users.
Inspired by these observations, in this work, we propose to leverage cold-start items to generate malicious item profiles for untargeted attacks in recommender systems.

\begin{figure}[t]
\centering
\subfigure[LastFM: NDCG@10]{\includegraphics[width=0.245\columnwidth]{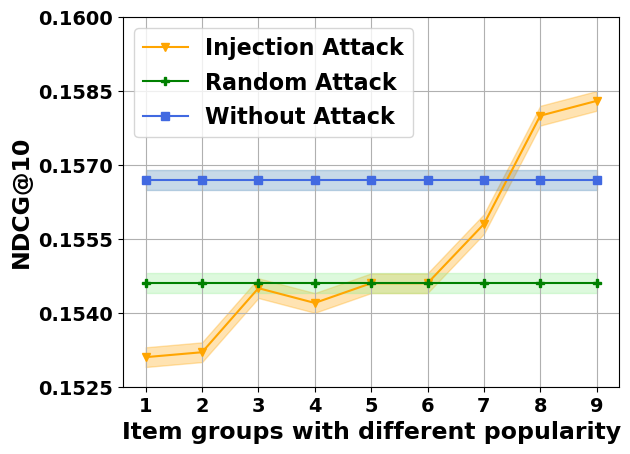}}
\subfigure[LastFM: RECALL@10]{\includegraphics[width=0.245\columnwidth]{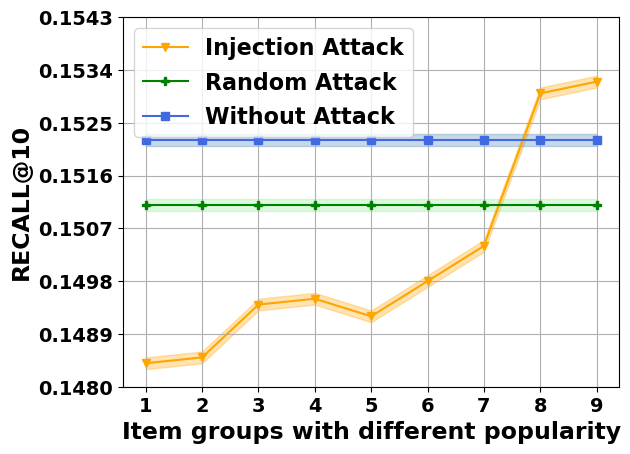}}
\subfigure[Ciao: NDCG@10]{\includegraphics[width=0.245\columnwidth]{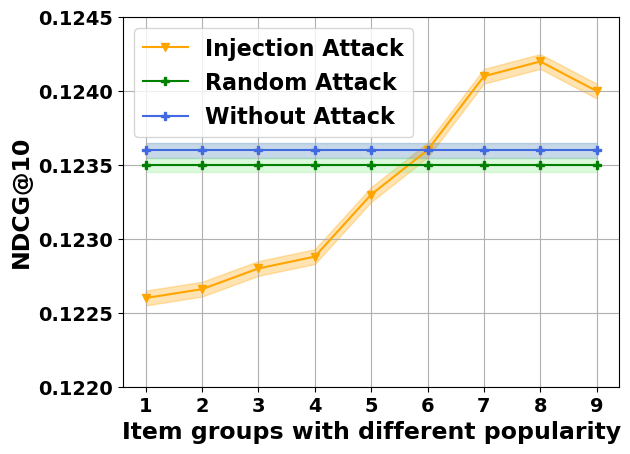}}
\subfigure[Ciao: RECALL@10]{\includegraphics[width=0.245\columnwidth]{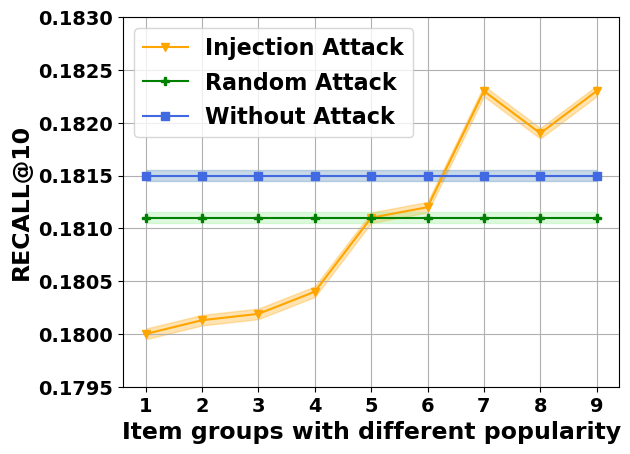}}
\vskip -0.12in
\caption{Comparing the impact of different filler items (i.e., item profile) based on popularity for untargeted attacks under a black-box social recommender system (e.g., GraphSAGE with social relations). Note that the items to the right of the x-axis are more popular.}
\label{fig:pre2}
\vskip -0.22in
\end{figure}

In order to validate the effectiveness of cold-start items for untargeted black-box attacks,  we conduct preliminary studies to examine the impact of utilizing filler items with varying popularity on attacks.
More specifically,  we divided items into nine groups based on their popularity (e.g., sales volume and page view) and compared their influence on social recommendation performance.  
Subsequently, we inject 10 fake users with item profiles from different popularity groups into the social recommender system. 
The experimental results are shown in Figure~\ref{fig:pre2}, where the items on the right side of x-axis indicate higher popularity. From the results, we can observe that injecting randomly constructed item profiles can lead to some degradation in recommendation performance.
Injecting items profiles with cold-start items into the recommender system leads to a more significant decrease in recommendation performance, especially the Top 10\% cold-start items, while injecting item profiles with popular items (e.g., the Top 20\% hottest items) can even improve recommendation performance. 
In other words, cold-start items are more damaging than popular items, indicating that selecting cold-start items as filler items tends to yield more effective untargeted attacks.

\begin{figure}[t]
\vskip -0.08in
\centering
\subfigure[LastFM: NDCG@5]{\includegraphics[width=0.24\columnwidth]{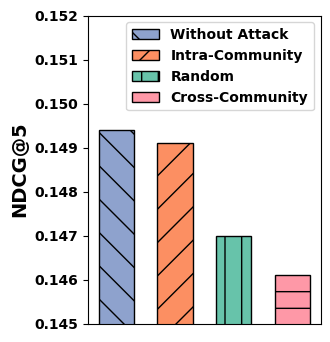}}
\subfigure[LastFM: NDCG@10]{\includegraphics[width=0.24\columnwidth]{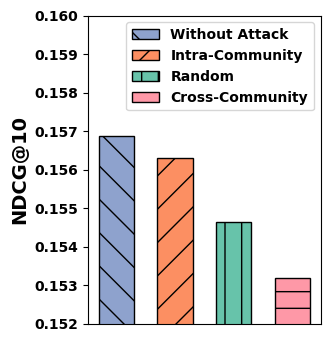}}
\subfigure[Ciao: NDCG@5]{\includegraphics[width=0.242\columnwidth]{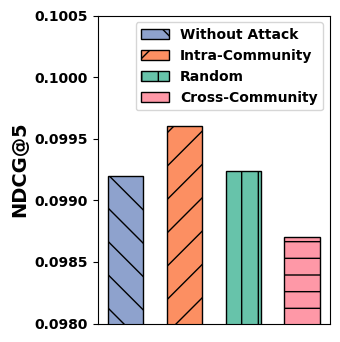}}
\subfigure[Ciao: NDCG@10]{\includegraphics[width=0.242\columnwidth]{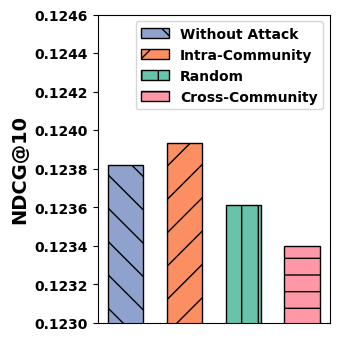}}
 \vskip -0.14in
\caption{Comparing the impact of different social connections (i.e.,  random connections,
 intra-community connections, and  cross-community connections) on untargeted attacks under the black-box social recommender system with injecting 10 fake users.}
\label{fig:pre1}
\vskip -0.3in
\end{figure}

\subsection{Cross-community Social Connections for Social Attacks} \label{communiy}

Different from the general attacking methods in recommender systems that aim to generate fake item profiles only, 
users' online social neighbors, which can significantly influence users' online behaviors, can be leveraged to enhance the attacking strategy in social recommender systems by generating fake social profiles.
Therefore, in this subsection, we aim to investigate how to establish social connections to conduct effective untargeted attacks on social recommender systems under the black-box setting.

In social networks, utilizing strong ties among users within a community can effectively enhance recommendation performance~\cite{gasparetti2021community,yin2014exploring}, as users within a community tend to have similar social characteristics and preferences towards items, even without any direct relationship.
In other words, the preferences of users within the same community are more similar than the outside ones.
Building upon this observation, connecting social neighbors among different communities might mislead social recommendations to capture users' preferences towards items. 
Thus, we introduce cross-community connections to generate fake social profiles for untargeted attacks in social recommender systems.

To verify the effectiveness of cross-community connections in attacking social recommendations, a preliminary study is conducted by injecting a few fake social profiles. 
Specifically,  we inject 10 fake users into a social recommender system (e.g., GraphSAGE~\cite{hamilton2017inductive} with social relations) to compare the impact of three different connection types on the recommendation performance: 1)  establishing random connections, 2) establishing intra-community connections, and 3) establishing connections across communities. 
To effectively divide the community, we employed the widely used unsupervised learning method Louvain algorithm~\cite{traag2019louvain} on the user-user social relations.
The experimental results on LastFM and Ciao datasets under the NDCG@$k$ ($k=5, 10$) metrics are shown in Figure~\ref{fig:pre1}. 
We can observe that cross-community connections on social profiles can lead to a more significant decrease in recommendation performance compared to intra-community and random connections.
In addition, generating social profiles by connecting users in the same communities (i.e., intra-community) on Ciao can result in a significant improvement in recommendation performance. 
These experimental findings suggest that constructing cross-community connections can more effectively degrade the recommendation performance and conduct more damaging attacks.

\section{Problem Statement}
In this section, we introduce some key notations and the research problem in this work.

\subsection{Notations and Definitions}
Let $U=\{u_1,u_2,\dots,u_n\}$ and $V=\{v_1,v_2,\dots,v_m\}$ represent the sets of users and items in the target social recommender system, where $n$ and $m$ are the sizes of users and items, respectively. 
We define user-item interactions as the matrix $\bm{Y} \in \mathbb{R}^{n\times m}$, 
where $y_{ij}=1$ if user $u_i$ has implicitly interacted with item $v_j$ (e.g., clicked and liked), and 0 otherwise.
An item profile towards a specific user $P_t=\{v_0,\dots,v_{t-1}\}$ consists of $t$ historical items that the user has interacted with in the history. 
Additionally, user-user social relations can be represented as the matrix $\bm{\mathcal{T}} \in \mathbb{R}^{n\times n}$, where $\bm{\mathcal{T}}_{ij}=1$ when user $u_j$ has a relation to user $u_i$, and 0 otherwise. In this work, the social recommendation function can be defined as:
\vskip -0.17in
\begin{equation}
\setlength{\abovedisplayskip}{2pt}
\setlength{\belowdisplayskip}{2pt}
y_{ui}=f(u,i),
\end{equation}
where $y_{ui}$ denotes the recommendation score for user $u$ and item $i$. Function $f$ is a composite function that considers both user-item preferences and social network information, aggregating them to produce a recommendation score. 
Generally, a recommendation model can then be trained using the Bayesian Personalized Ranking (BPR) loss function ~\cite{rendle2012bpr}:
\begin{equation}
\setlength{\abovedisplayskip}{2pt}
\setlength{\belowdisplayskip}{2pt}
\mathcal{L}_{rec} = \sum_{(u,i,j) \in \bm{\overline{Y}}}-\log(\sigma(y_{ui} - y_{uj}))+ \lambda||\theta^*||^2,
\end{equation}
where $\sigma$ is a sigmoid activation function, $\lambda$ are model specific regularization parameters to prevent overfitting, and $||\theta^*||^2$ denotes the $L_2$ normalization on the model parameter $\theta^*$.
$\bm{\overline{Y}} =\{(u, i, j)|(u, i)\in \bm{Y}, (u, j)\notin \bm{Y}\}$ consists of training triples $(u,i, j)$ that includes an observed user-item interaction and an unobserved one. The function measures the difference in predicted scores between a positive item $i$ and a negative item $j$ for each user $u$.

\subsection{Goal of Untargeted Black-box Attacks in Social Recommendations}

\noindent \textbf{Attacker's Knowledge.} 
We focus on untargeted attacks on black-box social recommender systems, in which the attacker is unable to access any details of the target recommender systems (e.g., the model's architecture and parameters), and user's online behaviors (i.e., user-item interactions), due to the safety and privacy concerns. 
In most social networking platforms, such as Quora and Zhihu, users' social relations are publicly available to access for connecting online friends with similar interests.  
In addition, these platforms also provide users with access to items' side information, including item titles, categories,  sales volume,  consumers' comments, etc. 
The availability of such details aids users in exploring and engaging with relevant content that aligns with their preferences and needs.

\noindent \textbf{Attacker's Goal.} 
In this work, the goal of untargeted attacks for social recommendations under the black-box setting is to degrade the recommendation performance by generating a set of carefully crafted fake users $U^F=\{u_{m+i}\}^\Delta_{i=1}$, where $\Delta={\Delta}_v+{\Delta}_u$ is the budget of item profiles and social profiles given to the attacker. 
These fake users $U^F$ consist of their \emph{item profiles} $P_t=\{v_0,\dots,v_{t-1}\}$ and \emph{social profiles} $S_t=\{...,u_j,...\}$, which are injected into user-item interactions $\bm{Y}$ and social graph $\bm{\mathcal{T}}$. Be aware that this can result in the target system having a set of malicious users $U'=U\cup U^F$, a polluted interaction matrix $\bm{\hat{Y}}\in ^{(\Delta_v+m)\times n}$ and a polluted social graph $\bm{\mathcal{\hat{T}}}\in ^{(\Delta_u+n)\times n}$. The attacker aims to degrade the overall performance of the recommendation model.
In summary, we can define the attack as follows:
\begin{equation}
\setlength{\abovedisplayskip}{3pt}
\setlength{\belowdisplayskip}{3pt}
    \begin{aligned}
    & \text{min}~\mathcal{L}_{attack}(f_{\theta_*}(\bm{\hat{Y}},\bm{\hat{\mathcal{T}}})) \\
     \text{s.t.} \ \
     {\theta_*} = \text{argm}
    & \text{in}_{\theta}(\mathcal{L}_{rec}(f(\bm{Y},\bm{\mathcal{T}}))+\mathcal{L}_{rec}(f(\bm{\hat{Y}},\bm{\hat{\mathcal{T}}}), \\
     ||\bm{\hat{Y}}&-\bm{Y}||\leq{\Delta}_v,||\bm{\hat{\mathcal{T}}}-\bm{\mathcal{T}}||\leq{\Delta}_u,
    \end{aligned}
\end{equation}
where $\mathcal{L}_{attack}$ can be expressed as $-\mathcal{L}_{rec}$.

%% file: sections/model-V2.tex
\section{The Proposed Method}
\label{sec:methodlogy}

\begin{figure*}[tb]
 \centering
\subfigure{\includegraphics[width=0.998\linewidth]{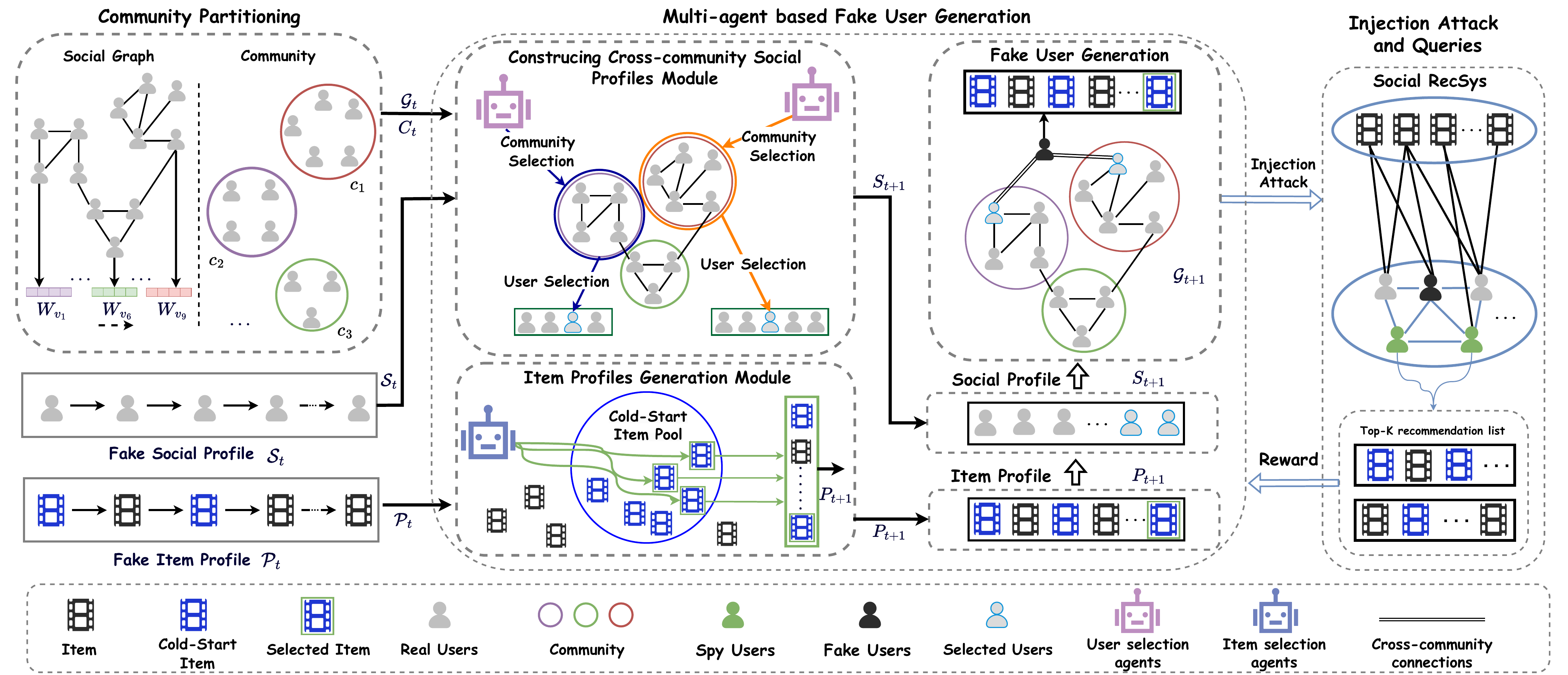}}
 \vskip -0.13in
\caption{An overview of the proposed framework. It contains three key components: community partitioning, multi-agent based fake user generation, and injection attack and queries in the target recommender system.}
\label{fig:framework}
\end{figure*}

In this section, we first present an overview of our proposed framework and provide an introduction to the attack environment. Then, we provide the details of each component of our proposed framework MultiAttack.

\subsection{An Overview of the Proposed Framework}
In order to perform an untargeted attack on the social recommendations under the black-box setting,
the attacker needs to generate item profiles and social profiles simultaneously. 
It's worth mentioning that the coordination and communication among item and social profiles generation are essential for achieving effective attacks in social recommendations, due to the intricate interconnections between item profiles and social profiles. 
However, it is challenging to utilize vanilla reinforcement learning with a single agent to maximize the overall reward in a complex environment~\cite{bucsoniu2010multi,zhang2021multi}. 
To address this challenge, a multi-agent reinforcement learning  based attacking framework  (\textbf{MultiAttack}) is introduced to train multiple agents to learn their own attacking policy in a decentralized manner, where these agents need to communicate with a central server and collaborate with others to take actions for optimal attacking strategy.  
More specifically, 
we develop two agents to learn the strategies for generating social profiles and another agent to learn the strategy for generating item profiles. 
The architecture of our proposed attack framework MultiAttack is illustrated in Figure ~\ref{fig:framework}, which consists of three main components: \emph{1) Community Partitioning}, \emph{2) Multi-agent Based Fake User Generation}, and \emph{3) Injection Attack and Queries}. 

The first component aims to partition the community on user-user social relations, so as to construct cross-community connections for the generation of social profiles.  
After determining the communities, the second component is proposed to craft social profiles and item profiles. 
Particularly,  a multi-agent based fake user generation component is developed to address the challenge of the coordination of social profiles and item profiles, which consists of two modules: \emph{constructing cross-community social profiles module} and \emph{item profiles generation module}. 
Meanwhile, a decentralized Actor-Critic architecture is proposed to guide these two modules to coordinate fake user generation.  
Once the second component generates a fake user with the item and social profiles, the goal of the third component is to attack and query the target social recommender system, and obtain a reward from the Top-$k$ recommendations list of spy users for optimizing the overall attacking framework. 
In the next subsection, we first provide an overview of the attack environment in our proposed MultiAttack framework.

\subsection{Attacking Environment Overview}
In order to perform effective attacks, we model the framework for attacking black-box recommender systems as a Decentralized Partially Observable Markov Decision Process (DEC-POMDP)~\cite{zhang2021multi,zhao2018deep},
in which one agent generates item profiles and the other generates social profiles for fake users.
The attacker injects the well-designed fake users into the environment and interacts with the environment (the black-box social recommender system) to maximize the expected cumulative reward from the environment.
Formally, we can define the DEC-POMDP as $\{\mathcal{S},\mathcal{A},\mathcal{O},\mathcal{P},\mathcal{R},\mathcal{D},\gamma \}$.

\begin{itemize}[leftmargin=*] 
\item \textbf{State Space $\mathcal{S}$:} 
The state $s_t\in \mathcal{S}$ is defined as the adjacency matrix of the social network with the item click sequence. Each agent can observe part of the state. In general, the state can be represented as a concatenation of the fake user's item profile and the fake user's social profiles, i.e., $P_t\cup S_t$.

\item \textbf{Action Space $\mathcal{A}$:} 
The action space is defined as $\mathcal{A}^s,\mathcal{A}^p \in \mathcal{A}$. 
The action $a_t=(a^s_t, a^p_t)\in \mathcal{A}$, where $a^s_t$ and $a^p_t$ select users and items to generate social profiles and item profiles, respectively. 
The action $a^s_t=(a^{c}_t, a^{u}_t)\in \mathcal{A}^s$ is to determine two actions to generate social profiles for connecting different communities, where the first action $a^{c}_t$ is to determine a specific community as the user candidates pool $C_t$, and another one $a^{u}_t$ is to pick one user from $C_t$ for generating social profiles. 
Note that two agents are developed to determine their action $a^{c}_t$ with the restriction to be different for building cross-community connections. 
The action $a^p_t=(a^{item}_t)\in \mathcal{A}^p$ is to pick one item from the item pool for generating item profiles. 

\item \textbf{Observation $\mathcal{O}$:} The observation $o^i_t=O(s_t; i)$ is the local observation for agent $i$ at global state $s$ at time step $t$, i.e., the agent generating social relations can observe the adjacency matrix state $s^s_t$, and the agent generating item profiles can observe the item click sequence state $s^p_t$. 

\item \textbf{Transition probability $\mathcal{P}$:} Transition probability $p(s_{t+1} | s_t, a_t)$ $\in \mathcal{P}$ defines the probability of state transition from the current $s_t$ to the next state $s_{t+1}$ when the agents take actions $a_t$.

\item \textbf{Reward $\mathcal{R}$:}  Once the generated social profiles and item profiles are injected into the $s_t$ state recommender system, the recommender system will provide the current reward (i.e., Top-$k$ recommended items on spy users) via a query~\cite{fan2021attacking}, in which cold-start items will be used to evaluate the effectiveness for untargeted attacks.
More detail about the reward definition can be found in Section~\ref{reward}.

\item  \textbf{Set of Agents $\mathcal{D}$:} 
The set of agents $\mathcal{D}=\{1,...,n\}$ consists of $n$ agents. Here $n$ is the total number of agents including agents that select items and generate social profiles. 
In this paper, we develop both three-agent and two-agent versions.

\item  \textbf{Discount factor $\gamma$:} The discount factor $\gamma \in [0,1]$ essentially determines the relative importance of immediate versus future rewards. A value of $\gamma=0$ means that the agent only considers the immediate reward, while a value of $\gamma=1$ means that the agent values all rewards equally, regardless of when they are received. Values of $\gamma$ between 0 and 1 allow the agent to balance the importance of immediate and future rewards.
\end{itemize}

\subsection{Community Partitioning}
Based on our preliminary studies in Section~\ref{communiy}, which indicate that cross-community connections can conduct more effective attacks in reducing recommendation performance, we aim to guide multiple agents in generating more malicious social profiles for fake users by incorporating a community partitioning strategy.
As illustrated in Figure~\ref{fig:framework}, to partition communities in large-scale social networks more efficiently, 
a representative clustering method K-means is used to perform community detection in an unsupervised learning way. 
More specifically, we only take advantage of the structural information by using network embedding learning techniques to learn nodes' representations. 
Hence, to obtain effective users' representations for clustering,  DeepWalk~\cite{perozzi2014deepwalk} is proposed to capture the local community structure by traversing nodes within the same community more frequently. The optimization objective can be defined as follows: 
\begin{equation}\label{eq:deepwalk}
\setlength{\abovedisplayskip}{2pt}
\setlength{\belowdisplayskip}{3pt}
\max \frac{1}{T} \sum_{t=1}^T \sum_{u_{t'} \in c(u_t)} \log p(u_{t'}|u_t),
\end{equation}
where $T$ is the total number of nodes in the random walk sequences. The probability $p(u_{t'}|u_t)$ is computed using the softmax function:
\begin{equation}\label{eq:softmax}
\setlength{\abovedisplayskip}{3pt}
\setlength{\belowdisplayskip}{3pt}
p(u_{t'}|u_t) = \frac{\exp\left(\mathbf{v}_{u_{t'}}^{\top} \mathbf{v}_{u_t}\right)}{\sum_{u=1}^{|U|} \exp\left(\mathbf{v}_u^{\top} \mathbf{v}_{u_t}\right)},
\end{equation}
where $\mathbf{v}_{u_t}$ and $\mathbf{v}_{u_{t'}}$ are the latent representations (embeddings) of users $u_t$ and $u_{t'}$, respectively, and $|U|$ is the number of users in the social network. 

It is worth noting that determining the appropriate number of clusters is crucial for the effectiveness of subsequent K-means to facilitate community clustering. 
Thus, we use Louvain's algorithm~\cite{blondel2008fast}, a prevalent unsupervised community detection method rooted in the optimization of modularity, to identify the number of clusters for community partitioning.

%% file: sections/model-V2-part2.tex
\begin{figure}[tb]
 \centering
\subfigure{\includegraphics[width=0.56\linewidth]{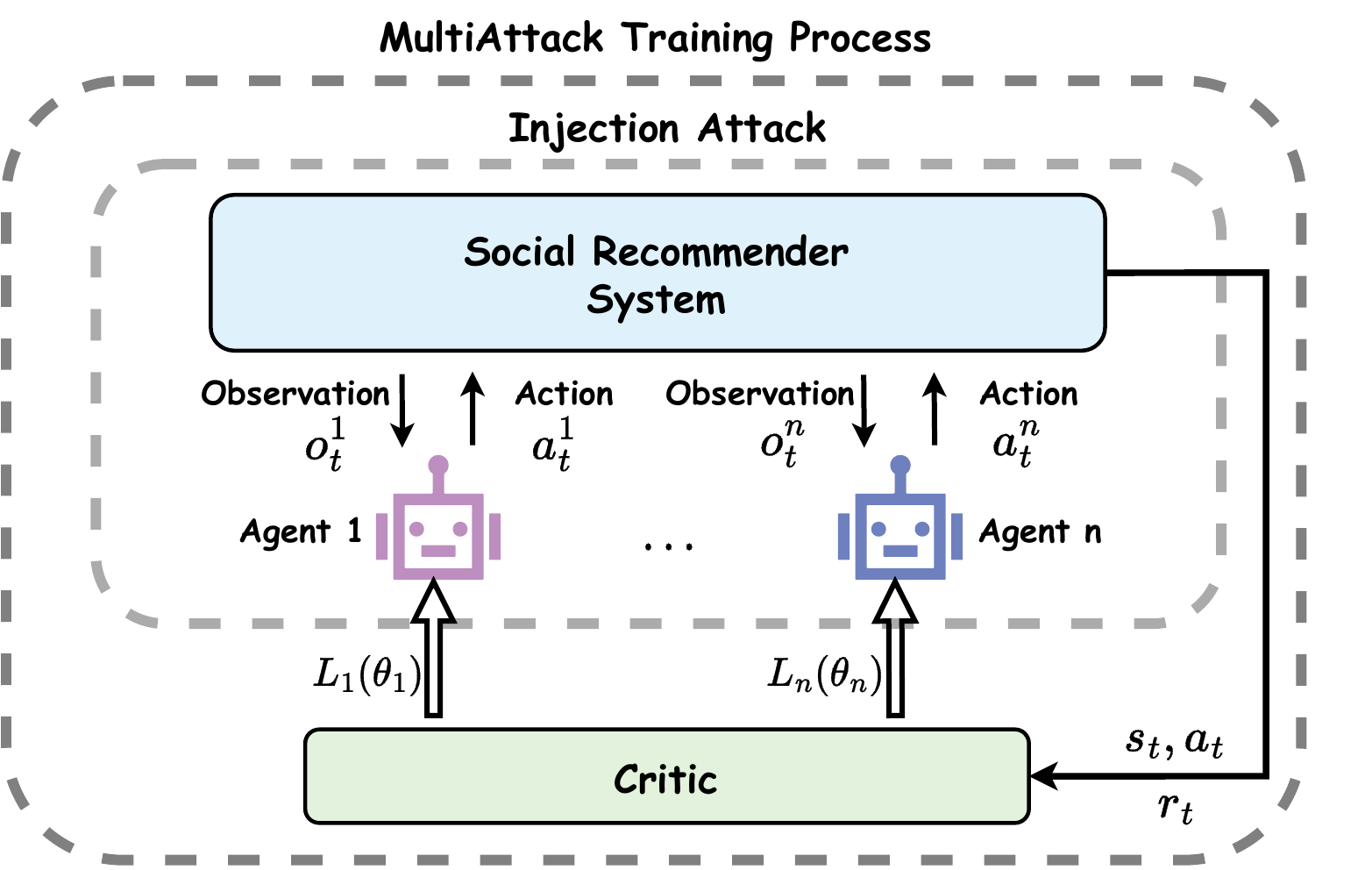}}
\caption{The decentralized Actor-Critic (A2C) based multi-agent reinforcement learning framework for attacking social recommendations.}
\label{fig:RL_training}
\vskip -0.2in
\end{figure}
\subsection{Multi-agent based Fake User Generation}
 After partitioning the community, the multi-agent policy networks aim to coordinate the generation of item profiles and social profiles to degrade recommendation performance as illustrated in the second component of Figure~\ref{fig:framework}. 
 To improve the efficiency of the attack and reduce the learning cost in a large-scale discrete action space, we develop \emph{three}  different agents to generate item profiles (one agent) and social profiles (two agents).
 Specifically, we develop three agents, where two agents are responsible for selecting different communities and users through a hierarchical module in social profiles generation, 
 and another agent aims to generate fake item profiles for the fake user.
 We also design a two-agent strategy within the proposed multi-agent framework, which is discussed in detail in Section~\ref{baseline}.

 Meanwhile, in order to enhance the cooperation between the two modules, we introduced the Actor-Critic (A2C) architecture (discussed in detail in Section~\ref{model-training}), each agent integrates an actor network which is denoted as $\pi_{\theta^i}(a_{t}^i|o_{t}^i)$  parameterized with $\theta^i$. 
 As illustrated in Figure~\ref{fig:RL_training}, each agent performs an action based on the observation $o^i_{t}$ from the social recommender system. For all agents, there is a centralized critic $V_{\omega}(s^1_{t}, \dots, s^n_{t})$ which is parameterized with $\omega$ to evaluate the action of each agent by the reward obtained from the social recommender system. With the guidance of the critic, each agent can learn an attack strategy to maximize the global reward, i.e. reduce the model recommendation performance.
  Next, we present each module in detail.

\subsubsection{\textbf{Constructing Cross-community Social Profiles Module}}
In order to effectively establish cross-community social profiles for fake users, two agents are developed in the social graph, where each agent's action can be defined into two steps: (1) community selection, and (2) user selection within the chosen community. 
Two agents act concurrently, with each agent selecting a distinct community and user. The joint action space for both agents consists of community-user pairs, and for each agent, the actor policy can be formulated as:
\centerline{
$\pi(a_t|o_t) = \pi_\theta^c(c_t|o_t) \cdot \pi_\theta^u(u_t|o_t, c_t),$
}
\noindent where $\pi_\theta^c(c_t|o_t)$ and $\pi_\theta^u(u_t|o_t, c_t)$ represent the policies for community selection and user selection within the community, respectively. 
It is worth noting that to construct cross-community user connections, we restrict the first action of the two agents to select different communities. 
Formally, the community selection policy can be represented through a neural network, where the input is the observation $o_t$, and the output is the distribution over community selections, denoted by:
\begin{equation}
\setlength{\abovedisplayskip}{3pt}
\setlength{\belowdisplayskip}{3pt}
\pi_\theta^c(c_t|o_t) = \text{Softmax}(\mathbf{W}_{\text{C},2}\text{ReLU}(\mathbf{W}_{\text{C},1}\mathbf{o}_{t})+\mathbf{b}_{\text{C}}),
\end{equation}
where $\theta=\{\mathbf{W}_{\text{C},1}, \mathbf{W}_{\text{C},2} \}$ are trainable parameters, and $\mathbf{b}_{\text{C}}$ is the bias. Then with the adapted advantage function, the hierarchical actor policy updates using the Policy Gradient (PG) algorithm~\cite{zhang2021multi}:
\vskip -0.15in
\begin{align}
A_{\omega}^c(o_t) &= \sum_{j=0}^{T-t}\gamma^{j} r^c_{t+j} - V_{\omega}(o_t), \\
\nabla_\theta J(\theta^c) &= \mathbb{E}_{\tau \sim \pi}\left[\sum_{t=0}^{|\tau|} \nabla_\theta \log \pi_\theta^c(c_t|o_t) A_{\omega}^c(o_t)\right].
\end{align}

The user selection policy follows a similar representation, where the input is the concatenation of the observation $o_t$ and the selected community $c_t$, and the output is the distribution over user selections within the selected community:
\begin{align}
\pi_\theta^u(u_t|o_t, c_t) &= \text{Softmax}(\mathbf{W}_{\text{U},2}\text{ReLU}(\mathbf{W}_{\text{U},1}\mathbf{x}_{t})+\mathbf{b}_{\text{U}}), \\
\mathbf{x}_t &= \text{Concat}(o_t, c_t),
\end{align}
where $\theta=\{\mathbf{W}_{\text{U},1}$, $\mathbf{W}_{\text{U},2}\}$ are trainable parameters, and $\mathbf{b}_{\text{U}}$ is the bias. The policy can be updated as:
\vskip -0.16in
\begin{align}
A_{\omega}^u(o_t, c_t) &= \sum_{j=0}^{T-t}\gamma^{j} r^u_{t+j} - V_{\omega}(o_t, c_t), \\
\nabla_\theta J(\theta^u) &= \mathbb{E}_{\tau \sim \pi}\left[\sum_{t=0}^{|\tau|} \nabla_\theta \log \pi_\theta^u(u_t|o_t, c_t) A_{\omega}^u(o_t, c_t)\right].
\end{align}

In summary, by employing constructing cross-community social profiles module with two agents working in parallel, each selecting users from a different community, we establish cross-community social connections to generate social profiles efficiently.

\subsubsection{\textbf{Item Profiles Generation Module}}
This module aims to select items to form fake item profiles for an untargeted attack. 
As detailed in preliminary studies of Section~\ref{cold-start}, injecting item profiles with cold-start items into the recommender system can lead to a more significant decrease in recommendation performance for untargeted attacks, especially the Top 10\% cold-start items.
Thus, to further reduce the action space, this module sample the Top $m\%$ of cold-start items as the item pool to generate item profiles, instead of all items. We determine cold-start items based on their popularity (e.g., sales volume and page view).
As shown in  Figure \ref{fig:framework}, this module first filters the item pool to pick out the cold-start items as candidate items $\mathcal{A}^i_{t}$. 
An agent is then introduced to learn the item selection policy $\pi_{\phi}(a_{t}^p|o_{t}^p)$, where $o_{t}^p$ is the selected item sequence (i.e., item profiles being generated) $P_t$ at step $t$.
The agent injects an item into item profiles $P_t$ to obtain $P_{t+1}$ over the action $a_{t}^p$ based on the observation $o^p_{t}$ and candidate items $\mathcal{A}^i_{t}$.

In addition, in order to generate high-quality item profiles, we develop a clip mechanism to clip item profiles $P_t$. Specifically, when the length of the item profile $P_t$ reaches $T$-length ($T\leq{\Delta}_v$), we clip the duplicate items to obtain $P^{clip}_t$ as the item profile for the fake user.
It is worth mentioning that due to adopting greedy strategies, a vanilla A2C method cannot explore generating item profiles effectively as it tends to fall into local maximum reward~\cite{zhang2021multi,chen2022knowledge}.
Hence, a decentralized A2C architecture is introduced to effectively avoid local optima, in which the critic encourages agents to explore collaborative strategies rather than optimizing individual item and social profile generation strategies.

\begin{figure}[tb]
\centering
{\subfigure[Top-10 List]
{\includegraphics[width=0.32\linewidth]{{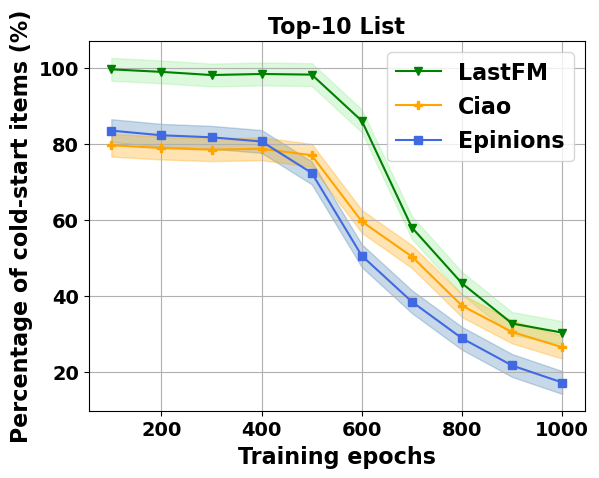}}}}
{\subfigure[Top-20 List]
{\includegraphics[width=0.32\linewidth]{{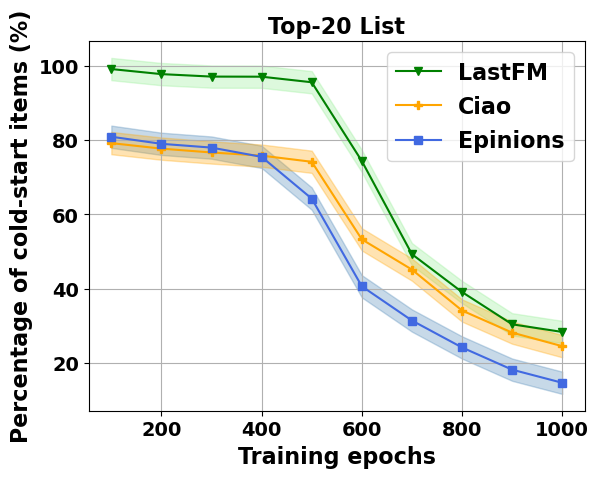}}}}
\vskip -0.14in
\caption{Hit ratio of cold-start items in Top-10 and 20 lists for users with different training epochs on three datasets on social recommender systems (i.e., GraphSAGE with social network).}
\label{fig:HR}
\vskip -0.15in
\end{figure}

\subsection{Injection Attack and Queries}\label{reward}
In order to attack the target social recommendation under the black-box setting, we need to inject the well-designed fake item profiles $P_{t}$ and social profiles $S_t$ into the system and perform a query to get reward feedback for optimizing the proposed multi-agent reinforcement learning based attacking framework, as illustrated in the last component of Figure~\ref{fig:framework}. 
Inspired by the general black-box attacking methods (i.e., CopyAttack and KGAttack) in recommender systems~\cite{fan2021attacking,chen2022knowledge}, we establish a set of spy users in the target recommender system to execute queries and obtain rewards for updating attacking policies.
However, different from the existing targeted attacks which can easily obtain the reward from the hit ratio of target items in the Top-$k$ recommendation list,  it is non-trivial to estimate the effectiveness of the generative fake user profiles (i.e., item profiles and social profiles) for \emph{untargetted attacks} under the black-box setting.
On the other hand, user-item interactions usually follow a long-tailed distribution~\cite{abdollahpouri2017controlling,brynjolfsson2011goodbye}, where a few popular items account for a large percentage of overall interactions, while a large number of cold-start items are rarely interacted with by users. 
In other words, most users prefer items with higher popularity to cold-start items, indicating that recommending more cold-start items to users is more likely to harm the overall recommendation performance.

To demonstrate this observation, we conducted a preliminary study on the social recommender system (GraphSAGE with social network) using LastFM, Ciao and Epinions datasets,  as illustrated in Figure~\ref{fig:HR}. 
During the training process in Top-$k$ recommended lists (i.e., $k=10, 20$), 
the performance of the model increases continuously, while the number of recommended cold-start items decreases. Notably, after 800 epochs of training, the hit ratio of cold-start items drops to less than 30\%. 
The experimental results indicate that recommending more cold-start items is harmful to recommendation performance.
Motivated by this observation, we define the hit ratio of cold-start items appearing in the recommendation list as rewards:
\begin{align}
r(s_t,a_t) =\frac{1}{|\hat{U}|}\sum^{|\hat{U}|}_{i=1} \text{HR}(\hat{u}_i, v^*, k),\\
\text{HR}(\hat{u}_i, v^*, k) =\frac{v_{\hat{u}}\cap v^*}{k},
\end{align}
where $\hat{u}_i\in\hat{U}$ is the spy user, $v^*$ is the set of cold-start items and $v_{\hat{u}}$ is the spy user’s Top-$k$ recommendation list. $\text{HR}(\hat{u}_i, v^*, k)$ returns the hit rate of the set $v^*$ in the Top-$k$ list of the spy user $\hat{u}_i$, i.e., the hit ratio of cold-start items in the spy user's Top-$k$ recommended items $v_{\hat{u}}$. Once the fake users' item profiles and social profiles have been generated, we perform injection and queries.

\subsection{Model Optimization}\label{model-training}
In this subsection, we present the optimization process of our proposed MultiAttack, including decentralized agent training and critic network updates. 
Without loss of generality, Multi-Agent Proximal Policy Optimization (MAPPO) is introduced to train the attack strategies~\cite{wu2021coordinated, yu2022surprising}. The training process of the proposed MultiAttack is summarized in Algorithm \ref{algo}.

\subsubsection{\textbf{Decentralized Agent Training}}
The optimization objective for all agents is developed to focus on maximizing the global reward, which can be formulated as:
\begin{equation}\label{obj}
\setlength{\abovedisplayskip}{1pt}
\setlength{\belowdisplayskip}{1pt}
    \max _{\pi} \mathbb{E}_{\tau \sim \pi}[R(\tau)], \text { where } R(\tau)=\sum_{t=0}^{|\tau|} \gamma^{t} r\left(s_{t}, A_{t}\right),
\end{equation}
where $\pi$ denotes the joint policy of all agents, and $\tau = (s_0, A_0,..., s_{T-1},$ $ A_{T-1})$ represents a trajectory of length $T$ that is generated based on the policy, and the trajectory consists of multiple state-action pairs. $A_t=(a^1_t,...,a^n_t)$ is the joint action at time step $t$. The expected value function from any state, $V^{\pi_{\theta}}(s_t)$, is represented as $\mathbb{E}{\pi_{\theta}}[\sum^{T-1}_{l=t} \gamma^{l-t} r_l]$, with $\theta$ denoting the policy parameters and $\gamma$ representing the discount factor. 

To effectively optimize the coordination of item profiles and social profiles, we adopt Actor-Critic (A2C)  framework~\cite{mnih2016asynchronous,wu2021coordinated}. As shown in Figure~\ref{fig:RL_training}, the attacker aims to train an adaptive attack policy that relies on multiple decentralized actors working in unison with a centralized critic. Specifically, each agent trains an optimal policy $\pi_{\theta^i} (a_i|o_i) $ parameterized with $\theta^i$ to produce an action $a_i$ from the local observation $o_i$, to maximize the discounted accumulated reward. 
These actor networks are then updated using policy gradients.

\setlength{\textfloatsep}{3pt}
\begin{algorithm}[tb]

\caption{\textbf{MultiAttack}}
\begin{algorithmic}[1]\label{algo}
\STATE Randomly initialize the Actor $\pi^c_{\theta}$, $\pi^u_{\theta}$, $\pi^v_{\theta}$ and Critic $V_{\omega}$ with parameters $\theta$, $\phi$ and $\omega$
\STATE Initialize replay memory buffer $\mathcal{D}$
\STATE Set learning rate $\alpha$, empty memory buffer $\mathcal{D}$
\FOR{episode number $c$ in $[0,\Delta / N)$}
\FOR{fake user $i$ in $[m+cN+1,m+(c+1)N+1]$}
\STATE Initialize state $s_0$ and observation $o_0$ for all agents
\FOR{step $t$ in $[0,T-1]$}
    \STATE Partition community $C_t$
    \FORALL{agent $i$ in social agents}
        \STATE Select community $C^i_t$ from $C_t$ according to  $\pi^c_{\theta}$
        \STATE Select user $u^i_t$ according to $\pi^u_{\theta}$, community $C^i_t$
        \STATE Inject user $u^i_t$ into social profile $S_t$ 
        \STATE Obtain observation $o^i_{t+1}=\{o^i_t,u^i_t\}$
    \ENDFOR
    \STATE Generate cold-start item pool ${A}^i_{t}$
    \FORALL{agent $j$ in item agents}
        \STATE Select item $v^j_t$ according to $\pi^v_{\theta}$, item pool ${A}^i_{t}$
         \STATE Inject item $v^j_t$ into item profile $P_t$
         \STATE Clip $P_t$ to obtain $P^{clip}_t$
        \STATE Obtain observation $o^j_{t+1}=\{o^j_t,v^j_t\}$
    \ENDFOR
    \STATE Obtain state $s_{t+1}=\{...,o^i_{t+1},...,o^j_{t+1},...\}$ and reward $r_t$
    \STATE $\tau+=[s_t,o^i_t,u^i_t,v^j_t,o^j_t,r_t,s_{t+1},o^i_{t+1},o^j_{t+1}]$
\ENDFOR
    \STATE Compute reward-to-go $\hat{R_t}$ on $\tau$ 
    \STATE Compute advantage estimate $\hat{A_t}$ on $\tau$ 
    \STATE Normalize $\hat{R_t}$ and $\hat{A_t}$ with PopArt
    \STATE Add $\tau$ into memory buffer $D$
\ENDFOR
\ENDFOR
\end{algorithmic}
\end{algorithm}

\subsubsection{\textbf{Critic Network Update}}
In this subsection, we introduce the updating of the critic network in detail. Our framework introduces a centralized critic network, $V_{\omega}(o^1_{t}, \dots, o^n_{t})$, which is parameterized with $\omega$ to determine the suitability of the chosen actions given the current state. 
To facilitate agents to coordinate social profiles and item profiles generation in a sparse reward landscape, 
we adopt a collaborative approach where all agents receive the global reward~\cite{wang2023novel, gupta2017cooperative}. 
Furthermore, to ensure improved convergence properties and reduced variance, we compute the advantage value $A_{\omega}(o^i_{t})$, which is the actual critic judgment based on the critic network's output~\cite{mnih2016asynchronous}. 
The critic network is updated using the temporal difference (TD) approach~\cite{tesauro1995temporal,tesauro1991practical} by minimizing the following squared error:
\begin{equation}\label{Critic_error}
\setlength{\abovedisplayskip}{3pt}
\setlength{\belowdisplayskip}{3pt}
\mathcal{L_{\omega}} = \sum_i \sum_t \left(\sum_{j=0}^{T-t}\gamma^{j} R_{t+j} - V_{\omega}(o^1_{t}, \dots, o^n_{t})\right)^2,
\end{equation}
where $R_{t+j}$ represents the shared global reward obtained by the agents at time step $t$. By minimizing this squared error, the critic network's parameters are updated to provide more accurate impact estimations of the agent's actions on the global rewards. 
In the end, guided by the judgments of the critic network, the actor networks adjust their parameters to enhance the attack performance, leading to improved actions in subsequent iterations. The number of iterations of parameter updates depends on the epoch length $K$, which influences the optimization strategy and value function.

\subsection{\textbf{Comparison of MultiAttack and Single-Agent Attack}}
In this subsection, we analyze the primary differences between MultiAttack and previous single-agent attack approaches~\cite{song2020poisonrec,fan2021attacking,chen2022knowledge}. \emph{Firstly}, MultiAttack employs a decentralized multi-agent reinforcement learning method to learn attack strategy. For single-agent attack methods, they only train one agent to learn to generate item profiles. It suggests that the single-agent attack method needs to additionally introduce other methods to generate social profiles. However, optimizing item and social profiles independently could easily lead to sub-optimal attacking performance on social recommendations due to their intricate interlinking. In contrast, MultiAttack trains multiple agents at the same time to collaboratively generate item profiles and social profiles. Therefore, our method can effectively coordinate the generation of social and item profiles for attacking black-box social recommendations. \emph{Secondly}, previous methods typically need to generate fake user profiles from a large-scale discrete action space, which makes it easy to generate low-quality user profiles. Instead, MultiAttack introduces multiple agents to generate item profiles and social profiles separately, which significantly reduces the time complexity. Additionally, MultiAttack also adopts two strategies to further reduce the agents’ action space so that they can learn more effective attack policies.

\subsection{\textbf{Time Complexity Analysis}}
In this subsection, we analyze the time complexity.
For general RL-based attack methods~\cite{chen2022knowledge,song2020poisonrec}, the time complexity of making the decision using reinforcement learning is typically linear to their action space. For example, let $U$ and $V$ represent the sets of users and items in the target social recommender system. As the agent selects items from the item pool to generate an item profile, the action space is $O(|V|)$. When generating social profiles, the agent needs to connect two different users and the action space of this operation is $O(|U|^2)$. Therefore for the vanilla single-agent RL-based black-box attack methods on social recommendations, the complexity of their action space is $O(|V|\times |U|^2)$. 

However, the number of items and users in real life can be in the millions. Therefore, we introduce decentralized multi-agent reinforcement learning to decrease the action space to reduce the time complexity. In our setting, MultiAttack adopts three agents to generate item profiles and social profiles. To further reduce the action space, we employ cold-start items and cross-community social connections. We can use $A$ and $C$ to represent the sets of cold-start items and community users, where $A\ll V$ and $C\ll U$. Therefore, for each agent in our framework, the complexities of their action space are $O(|A|)$, $O(n\times |C|)$ and $O(n\times |C|)$, respectively. Where $n$ is the number of communities. As each agent makes decisions individually based on their observations, the complexity of the overall framework for MultiAttack is $O(|A|)+O(n\times |C|)+O(n\times |C|)$. Therefore, the complexity is significantly less than typical black-box attack methods and is acceptable in practice.

%% file: sections/experiments-V2.tex
\section{Experiment}
\label{sec:Experiments}

In this section, we conduct extensive experiments to demonstrate the effectiveness of our proposed framework.
We first introduce the experimental settings, including the datasets and evaluation metrics employed. 
We then present the results of our approach and compare its performance with various baseline methods.
Finally, we study the impact of different components on our model and analyze the parameters used in our method to provide insights into the effectiveness of our approach.

 \subsection{\textbf{Experimental Settings}}

\subsubsection{\textbf{Datasets}}\label{datasets}
We use three real-world social datasets to evaluate the effectiveness of our proposed MultiAttack.
\textbf{LastFM}\footnote{http://millionsongdataset.com/lastfm/} is a dataset containing user-musician listening interactions from the Last.FM online music system. 
\textbf{Ciao}\footnote{https://www.cse.msu.edu/$\sim$tangjili/datasetcode/ciao} and \textbf{Epinions}\footnote{http://www.trustlet.org/epinions.html} are social datasets that record user-item interactions in the context of product recommendations in their respective online communities. The statistics of these three datasets are presented in Table~\ref{tab:datasets}.

\begin{table}[t]
  \centering
  \caption{Statistics of the datasets.}
    \scalebox{0.89}
{
    \begin{tabular}{c|c|c|c}
    \toprule
    \multirow{1}{*}{Datasets} & {LastFM} & {Ciao} & {Epinions} \\
    \midrule
User & 1,892 & 7,375 & 40,163   \\
Item  & 17,632 & 104,975 & 139,738   \\
Interaction & 92,834 & 226,307 & 664,824 \\
Density & 0.278\% & 0.0281\% & 0.0118\% \\
 \midrule
Social Relation & 25,434  & 111,781 & 487,183 \\
Density & 0.711\% & 0.2087\% & 0.0118\% \\
    \bottomrule
    \end{tabular}%
}
  \label{tab:datasets}%
\end{table}%

\subsubsection{\textbf{Baselines}}\label{baseline}

In the realm of social recommender system research, our work pioneers the investigation into untargeted attacks under the black-box setting. 
Thus,  to evaluate the effectiveness, we compare our proposed method with meticulously crafted a series of baselines:

\begin{itemize}[leftmargin=*] 
\item \textbf{ColdAttack}: 
This baseline randomly samples items from the cold-start item pool to construct the fake item profiles, and randomly samples a set of users to construct social profiles. 

\item \textbf{Popularity Perturbation (PopAttack)}~\cite{burke2005limited}: 
This baseline is an untargeted version of the explicit boosting in the bandwagon attack, where fake users interact with some popular items and randomly sample users to construct social profiles. 
We select the top $k$ (e.g., $k=10$) of the most popular items as candidate items.

\item \textbf{RandomAttack}: 
This baseline involves randomly sampling items from the entire item space $V$ to construct item profiles and randomly sampling users to construct social profiles.

\item \textbf{DegreeAttack}: 
This baseline randomly samples items from the entire item space $V$ to craft the fake item profiles and construct social profiles with users who have a high degree in the social network.  
To be specific, we select users with degree greater than $n$ (e.g., $n=20$) as candidate users.

\item \textbf{PoisonRec-Untargeted}~\cite{song2020poisonrec}:  
This baseline is an untargeted version of PoisonRec. 
It uses Proximal Policy Optimization (PPO)~\cite{schulman2017proximal} to generate item profiles and social connections. Specifically, it uses one agent to generate users' social profiles and item profiles. In order to conduct untargeted attacks, we set its rewards consistent with MultiAttack.

\item \textbf{MultiAttack-Social}: This method is a variant of MultiAttack, which only adopts \emph{constructing cross-community social profiles module} to generate fake user social profiles and randomly samples items from the entire item space $V$ to construct fake item profiles.

\item \textbf{Double-Attack}: This method is a two agents version of MultiAttack.  
Specifically, the \emph{constructing cross-community social profiles module} adopt an agent to perform four actions: (1) select community $c^1_t$, (2) select user $u^1_t$ from community $c^1_t$, (3) select community $c^2_t$, (4) select user $u^2_t$ from community $c^2_t$. The
\emph{item profiles generation module} is the same as MultiAttack's.
\end{itemize}

\subsubsection{\textbf{Attacking Environment: Target Social Recommender System}}
\begin{itemize}[leftmargin=*]
\item \textbf{Evasion Attack} (Model Testing Stage). 
In this stage, the recommendation model’s parameters are fixed without any retraining on the polluted datasets. 
To conduct evasion attacks, we build a GraphSAGE based social recommender system. Specifically, we develop a 3-layer GraphSAGE~\cite{hamilton2017inductive,fan2019graph} which can learn user and item representations from both user-item interactions and social graphs for making recommendations.

\item \textbf{Poison Attack} (Model Training Stage). 
In this stage, the recommendation model's training data can be manipulated by injecting fake users, leading to the retraining of the target recommender system on the poisoned datasets. We conduct poison attacks on a \textbf{Black-box} social recommender system \textbf{SocialLGN} (i.e., incorporating social relations into \textbf{LightGCN}) ~\cite{liao2022sociallgn}. To validate the effectiveness of the attack framework against various black-box recommendation models, we also perform poisoning attacks on the traditional collaborative filtering-based social recommender system (i.e., SBPR~\cite{zhao2014leveraging}).
\end{itemize}

\subsubsection{\textbf{Evaluation Metrics}}

 To evaluate the effectiveness of our attack on social recommender systems, we employ three widely used evaluation metrics: Normalized Discounted Cumulative Gain (NDCG@$k$),  Recall@$k$ and Precision@$k$ ~\cite{Wang2019NeuralGC,He2020LightGCNSA}. We present results for $k$ equal to 5, 10 and 20.
Note that small values of these metrics indicate better attacking performance for untargeted attacks.

\subsubsection{Parameter Settings}
Our proposed framework is implemented on the basis of PyTorch. 
The profile length $T$ for fake users is set to 30. 
To obtain the reward, we randomly select 50 spy users to perform queries per 2 injections and select the Top 10\% cold-start items as items pool.
For all actor and critic networks, we set 4 layers for LastFM and 6 layers for Ciao and Epinions. 
The layer dimension for LastFM is 32 and 64 for Ciao and Epinions. The learning rate and discount factor for rewards are set to 0.0005 and 0.99, respectively. 
The generalized advantage estimation lambda is set to 0.95. 
We set the entropy term coefficient to 0.01 and the value loss coefficient to 0.5. The PPO clip parameter is set to 0.2. In addition, to get a better performance of the model, we use PopArt to normalize the reward~\cite{hessel2019multi}.
For LastFM, Ciao, and Epinions datasets, we set the budget for fake users to 2\%, 2\%, and 0.5\% of the number of normal users, respectively.

\begin{table*}[t]
\centering
  \caption{Performance comparison of different black-box attack methods on target inductive social recommender system. We use bold
fonts and underline to label the best performance and the best baseline performance, respectively. N@$k$, R@$k$ and P@$k$ denote
NDCG@$k$, RECALL@$k$ and PRECISON@$k$, respectively.}
    \scalebox{0.73}
{
    \begin{tabular}{c|c|c|c|c|c|c|c|c|c|c}
    \toprule
    \multirow{1}{*}{Dataset}  & \multicolumn{1}{c|}{Algorithms} & \multicolumn{1}{c|}{N@5} & \multicolumn{1}{c|}{N@10} & \multicolumn{1}{c|}{N@20} & \multicolumn{1}{c|}{R@5}& {R@10}  & \multicolumn{1}{c|}{R@20} & \multicolumn{1}{c|}{P@5} & \multicolumn{1}{c|}{P@10} & \multicolumn{1}{c}{P@20} \\
\midrule
    \multirow{10}{*}{\textbf{LastFM}}&  Without Attack & 0.1493  & 0.1567  & 0.2064 & 0.0825 &0.1522 & 0.2604  & 0.1483  & 0.1373  & 0.1191   \\
    &ColdAttack & 0.1446  & 0.1518  & 0.1990  & 0.0807 & 0.1491  & 0.2523  & 0.1433 & 0.1339 & 0.1148   \\
    &PopAttack& 0.1532  & 0.1586  & 0.2088  & 0.0846 &0.1541 & 0.2631 & 0.1509  & 0.1383 & 0.1201  \\
    &RandomAttack & 0.1407  & 0.1474  & 0.1967  & 0.0785 & 0.1438  & 0.2506  & 0.1389 & 0.1298 & 0.1141   \\
     \cmidrule{2-11}
    &DegreeAttack& 0.1409  & \underline{0.1472}  & \underline{0.1936}  & 0.0780 & \underline{0.1429} & \underline{0.2440} & 0.1384  & \underline{0.1290} & \underline{0.1117} \\
    \cmidrule{2-11}
    &  PoisonRec-Untargeted & \underline{0.1407}  & 0.1488  & 0.1975 & \underline{0.0780} &0.1460 & 0.2517  & \underline{0.1381}  & 0.1303  & 0.1139   \\
    \cmidrule{2-11}
    &MultiAttack-Social & 0.1217  & 0.1264  & 0.1669  & 0.0673 & 0.1228  & 0.2115  & 0.1160 &  0.1081 & 0.0940   \\
    &\textbf{DoubleAttack}& 0.1148  & 0.1186  & 0.1578  & 0.0640 &0.1170 & 0.2012 & 0.1146  & 0.1057& 0.0911  \\
    & \textbf{ MultiAttack} & \textbf{0.1127}  & \textbf{0.1184}  & \textbf{0.1567} & \textbf{0.0616} &\textbf{0.1155} & \textbf{0.1994}
  & \textbf{0.1104}  & \textbf{0.1045}  & \textbf{0.0906}   \\
    \cmidrule{2-11}
     & \textbf{Improvement} &\textbf{ 19.9\%} & \textbf{ 19.5\%}  &\textbf{ 19.1\%} & \textbf{ 21.0\%}  & \textbf{19.1\%} &\textbf{ 18.3\%} & \textbf{ 20.0\%} &\textbf{ 19.0\%}  & \textbf{ 18.9\% }\\

   \midrule
    \multirow{10}{*}{\textbf{Ciao}}&  Without Attack & 0.0992  & 0.1236  & 0.1551 & 0.1138 &0.1815 & 0.2753  & 0.0650  & 0.0559  & 0.0458   \\
    &ColdAttack & 0.0979  & 0.1231 & 0.1538  & 0.1116 & 0.1816  & 0.2734  & 0.0645 & 0.0556 & 0.0452   \\
    &PopAttack& 0.1001  & 0.1247  & 0.1567  & 0.1146 &0.1837 & 0.2792 & 0.0660  & 0.0563 & 0.0462  \\
    &RandomAttack & 0.0974  & 0.1219  & 0.1530  & 0.1122 & 0.1809  & 0.2736  & 0.0648 & 0.0555 & 0.0453   \\
     \cmidrule{2-11}
    &DegreeAttack& \underline{0.0936}  & \underline{0.1190}  & \underline{0.1507}  & \underline{0.1078} &\underline{0.1768} & \underline{0.2712} & \underline{0.0612}  & \underline{0.0539}& \underline{0.0445}  \\
    \cmidrule{2-11}
    &  PoisonRec-Untargeted & 0.0966  & 0.1221  & 0.1528 & 0.1080 &0.1786 & 0.2716 & 0.0635  & 0.0554  & 0.0450   \\
    \cmidrule{2-11}
    &MultiAttack-Social & 0.0875  & 0.1132  & 0.1448  & 0.0991 & 0.1699  & 0.2651  & 0.0586 & 0.0527 & 0.0440   \\
    &\textbf{DoubleAttack}& 0.0827  & 0.1066  & 0.1384 & 0.0952 &0.1617 &0.2571 & 0.0558  & 0.0497 & 0.0424  \\
    &  \textbf{MultiAttack} & \textbf{ 0.0823 } & \textbf{ 0.1061 }  & \textbf{ 0.1369 } & \textbf{ 0.0946 } &\textbf{ 0.1605 } &\textbf{ 0.2538 }  & \textbf{ 0.0557}  & \textbf{ 0.0497 }  & \textbf{ 0.0418 }   \\
    \cmidrule{2-11}
     & \textbf{Improvement} &\textbf{ 12.0\%} & \textbf{ 10.8\%}  &\textbf{ 9.1\%} & \textbf{ 12.2\%}  & \textbf{ 9.2\%} &\textbf{ 6.4\%} & \textbf{ 8.9\%} &\textbf{ 7.7\%}  & \textbf{ 6.1\% }\\
    
    \midrule
    \multirow{10}{*}{\textbf{Epinions}}&  Without Attack & 0.1458  & 0.1946  & 0.2419 & 0.2022 &0.3338 &0.4820  & 0.0764  & 0.0684  & 0.0547   \\
    &ColdAttack & 0.1192  & 0.1696  & 0.2189  & 0.1768 & 0.3109  & 0.4653  & 0.0623 & 0.0606 & 0.0508   \\
    &PopAttack& 0.1435  & 0.1934  & 0.2414  & 0.2031 &0.3370 & 0.4859 & 0.0751  & 0.0682& 0.0550  \\
    &RandomAttack & 0.1181  & 0.1678  & 0.2173 & 0.1760 & 0.3083  & 0.4638  & 0.0618 & 0.0598 & 0.0505   \\
     \cmidrule{2-11}
    &DegreeAttack& 0.1332  & 0.1826  & 0.2306  & 0.1909 &0.3227 &0.4731 & 0.0697  & 0.0646 & 0.0527  \\
    \cmidrule{2-11}
    &  PoisonRec-Untargeted &\underline{0.1161}  & \underline{0.1661}  & \underline{0.2155} & \underline{0.1733} & \underline{0.3069}  & \underline{0.4612} & \underline{0.0611}  & \underline{0.0595}  & \underline{0.0503}   \\
    \cmidrule{2-11}
    &MultiAttack-Social & 0.1303  & 0.1812 & 0.2301  & 0.1891 & 0.3256  & 0.4777  & 0.0681 &  0.0641 &  0.0530   \\
    &\textbf{DoubleAttack}& \textbf{0.1038}  & \textbf{0.1537}  & \textbf{0.2042}  & \textbf{0.1589} &0.2929 & 0.4516 & \textbf{0.0541}  & \textbf{0.0548}& 0.0480  \\
    &  \textbf{MultiAttack}& 0.1042  & 0.1542  & 0.2046 & 0.1594 & \textbf{0.2925} & \textbf{0.4510}  & 0.0542  & 0.0549  & \textbf{0.0480}   \\
    \cmidrule{2-11}
     & \textbf{Improvement} &\textbf{ 10.5\%} & \textbf{7.4\%}  &\textbf{ 5.2\%} & \textbf{8.3\%} & \textbf{ 4.6\% } & \textbf{ 2.2\%} &\textbf{ 11.4\%} & \textbf{ 7.8\%} &\textbf{ 4.5\%} \\

    \bottomrule
    \end{tabular}%
}
  \label{tab:results}%
\end{table*}%

\subsection{\textbf{Overall Attacking Performance Comparison}}
\subsubsection{\textbf{Evasion Attack}}
We first compare the attacking performance of different methods on an inductive GNN-based social recommender system under the black-box setting, as shown in Table ~\ref{tab:results}. 
From the comparison, we have the following main observations:

\begin{itemize}[leftmargin=*]
\item As naive heuristic methods, ColdAttack, PopAttack, and RandomAttack show limited progress against without attacks on LastFM and Ciao datasets. This is mainly due to they cannot generate effective social profiles. In contrast, the attack performance on black-box social recommender systems is significantly enhanced when social relations are incorporated, such as PoisonRec-Untargeted and DegreeAttack.

\item For larger datasets like Epinions, DegreeAttack becomes less effective due to the complexity of the involved social network. 
In addition, ColdAttack improves performance over PopAttack by about 5\% to 15\%, validating the advantage of selecting cold-start items to generate item profiles can perform more effective attacks.

\item MultiAttack-Social is less effective than MultiAttack, especially over 20\% lower on the Epinions dataset.
The results demonstrate that an effective attack requires coordinating the generation of social profiles and item profiles, highlighting the necessity for our framework to build multi-agent to collaborate.

\item Both MultiAttack and its variant, DoubleAttack, achieve superior performance over all baseline methods, 
with significant improvements on all three datasets, indicating the effectiveness of our proposed method in attacking social recommender systems. 
Furthermore, MultiAttack and DoubleAttack show a substantial improvement over PoisonRec-Untargeted across various dataset sizes, showing the effectiveness of our decentralized framework in coordinating the social and item profiles generations. 
On LastFM and Ciao datasets, MultiAttack outperforms DoubleAttack, indicating that the three agents can explore more effectively in most cases. However, on the Epinions dataset, DoubleAttack slightly outperforms MultiAttack, possibly due to the fact that the complex non-stationary environment increases the learning cost for the agents, making two agents more stable in this scenario.
\end{itemize}

\subsubsection{\textbf{Poison Attack}}
We evaluate the attacking performance under black-box social recommender systems (i.e., \textbf{SocialLGN ~\cite{liao2022sociallgn}}) on LastFM and Ciao datasets. The experimental results are shown in Table~\ref{tab:poison1}. We present results for NDCG@$k$, Recall@$k$, and Precision@$k$ metrics, with $k=$ 5 and 10.
We exclude the results of some baselines (e.g., PopAttack, RandomAttack, MultiAttack-Social, and DoubleAttack), since similar observations as Table~\ref{tab:results} can be made.
We can observe that our proposed method MultiAttack consistently outperforms all baselines on these two datasets, highlighting the effectiveness of our framework for generating high-quality social profiles and item profiles, and coordinating two profile generations. 
It's important to note that these experimental results demonstrate that our proposed attacking framework can transfer to various recommendation models including inductive and transductive GNN-based recommender systems.

\subsubsection{\textbf{Black-box Attack Analysis.}} We evaluate our attack method on different types of recommender systems to verify the effectiveness of attacks on black-box social recommender systems. By observing Table~\ref{tab:poison2} and Table~\ref{tab:poison1}, we can find that the MultiAttack framework can significantly degrade the recommendation performance of both traditional collaborative filtering-based methods (e.g., SBPR) and state-of-the-art GNNs-based methods (e.g., SocialLGN). These comparison results highlight the effectiveness and generalization capability of the MultiAttack under the black-box setting.

\begin{table}[t]
\centering
  \caption{Performance comparison of different black-box attack methods on LastFM and Ciao datasets across three metrics. N@$k$, R@$k$ and P@$k$ denote NDCG@$k$, RECALL@$k$ and PRECISON@$k$, respectively.}
    \scalebox{0.9}
{
    \begin{tabular}{c|c|c|c|c|c|c}
    \toprule
    \multirow{2}[3]{*}{Dataset} & \multicolumn{6}{c}{LastFM} \\
\cmidrule{2-7}             
& \multicolumn{1}{c}{N@5} & \multicolumn{1}{c|}{N@10} & \multicolumn{1}{c}{R@5} & \multicolumn{1}{c|}{R@10} & \multicolumn{1}{c}{P@5} & \multicolumn{1}{c}{P@10} \\
\midrule
    Without Attack & 0.2946  & 0.2533  & 0.1350 & 0.1993 &0.2616 & 0.1947     \\
    ColdAttack & 0.2937  & 0.2511  & 0.1348  & 0.1975 & 0.2616  & 0.1923    \\
    DegreeAttack & 0.2945  & 0.2526  & 0.1351  & 0.1985 &0.2617 & 0.1937    \\
    PoisonRec-Untargeted & 0.2935  & 0.2528  & 0.1339  & 0.1991 &0.2602 & 0.1942   \\
    \midrule
    \textbf{MultiAttack} & \textbf{0.2851}  & \textbf{0.2435}  & \textbf{0.1308}  & \textbf{0.1912} &\textbf{0.2556} & \textbf{0.1870}  \\
    \bottomrule
    \multirow{2}[3]{*}{Dataset} & \multicolumn{6}{c}{Ciao} \\
\cmidrule{2-7}             
& \multicolumn{1}{c}{N@5} & \multicolumn{1}{c|}{N@10} & \multicolumn{1}{c}{R@5} & \multicolumn{1}{c|}{R@10} & \multicolumn{1}{c}{P@5} & \multicolumn{1}{c}{P@10} \\
\midrule
    Without Attack & 0.0313  & 0.0327  & 0.0225 & 0.0344 &0.0249 & 0.0194     \\
    ColdAttack & 0.0308  & 0.0323  & 0.0223  &0.0339   & 0.0243  & 0.0191    \\
    DegreeAttack & 0.0306  & 0.0323  & 0.0220  & 0.0333 &0.0238 & 0.0192    \\
    PoisonRec-Untargeted & 0.0306  &  0.0325  & 0.0222  & 0.0343 &0.0241 & 0.0194   \\
    \midrule
    \textbf{MultiAttack} & \textbf{0.0280}  & \textbf{0.0297}  & \textbf{0.0195}  & \textbf{0.0308} &\textbf{0.0221} & \textbf{0.0177}  \\
    
    \bottomrule
    \end{tabular}%
}
  \label{tab:poison1}%
  \vskip 0.1in
\end{table}%

\begin{table}[t]
\centering
  \caption{Performance comparison of different black-box attack methods on target recommender system (SBPR) across three metrics. N@$k$, R@$k$ and P@$k$ denote NDCG@$k$, RECALL@$k$ and PRECISON@$k$, respectively.}
    \scalebox{0.9}
{
    \begin{tabular}{c|c|c|c|c|c|c}
    \toprule
    \multirow{2}[3]{*}{Dataset} & \multicolumn{6}{c}{LastFM} \\
\cmidrule{2-7}             
& \multicolumn{1}{c}{N@5} & \multicolumn{1}{c|}{N@10} & \multicolumn{1}{c}{R@5} & \multicolumn{1}{c|}{R@10} & \multicolumn{1}{c}{P@5} & \multicolumn{1}{c}{P@10} \\
\midrule
    Without Attack & 0.2084  & 0.1998  & 0.0738 & 0.1323 &0.1571 & 0.1409     \\
    ColdAttack &0.2084   &0.2023   &0.0738   &0.1329  &0.1557   &0.1408     \\
    DegreeAttack &0.2111   &0.2061   &0.0730   &0.1340  &0.1555   & 0.1428     \\
    PoisonRec-Untargeted &0.2059   &0.2026   &0.0746   &0.1311  &0.1562   &0.1391     \\
    \midrule
    \textbf{MultiAttack} &\textbf{0.1990}  &\textbf{0.1935}   &\textbf{0.0702}   &\textbf{0.1283}  &\textbf{0.1479}   &\textbf{0.1364}     \\
    \bottomrule 
    \end{tabular}%
}
  \label{tab:poison2}%
  \vskip 0.1in
\end{table}%

 \begin{figure*}[t]
 \centering
 \centering
{\subfigure[LastFM]
{\includegraphics[width=0.15\linewidth]{{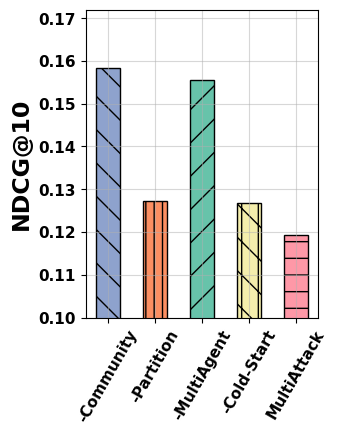}}}}
{\subfigure[Ciao]
{\includegraphics[width=0.155\linewidth]{{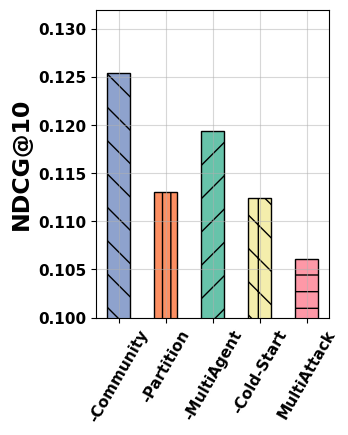}}}}
{\subfigure[Epinions]
{\includegraphics[width=0.15\linewidth]{{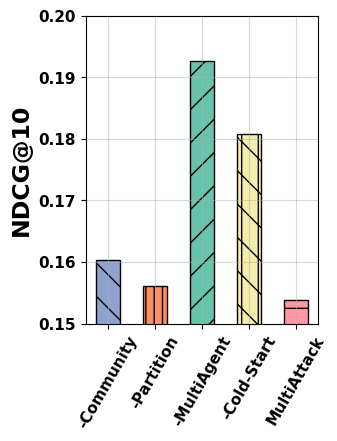}}}}
{\subfigure[LastFM]
{\includegraphics[width=0.15\linewidth]{{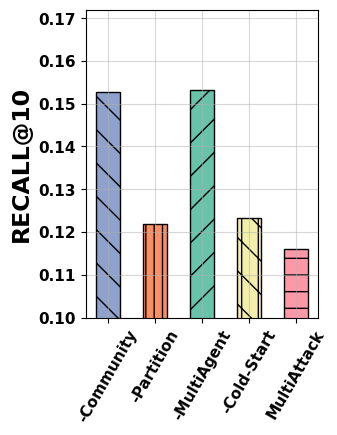}}}}
{\subfigure[Ciao]
{\includegraphics[width=0.155\linewidth]{{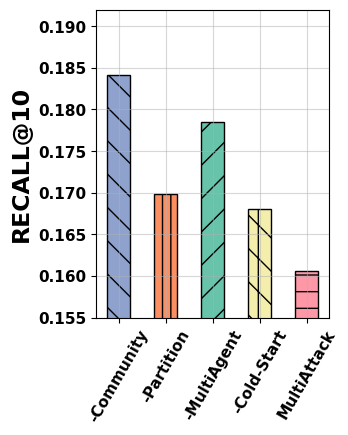}}}}
{\subfigure[Epinions]
{\includegraphics[width=0.15\linewidth]{{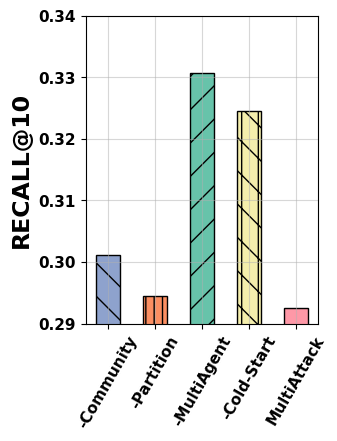}}}}
\vskip -0.12in 
 \caption{Comparison among MultiAttack and its variants on three datasets across metrics NDCG@10 and RECALL@10 for inductive social recommender system. }\label{fig:ablation}
 \end{figure*}

\subsection{\textbf{Ablation Study}}
In order to measure the impact of each component of our proposed framework MultiAttack, we compare its performance against several of its ablated versions.  Specifically, we consider the following ablations:
(1) MultiAttack (-Community): This method removes the cross-community connection aspect, allowing agents to build social profiles freely with all users.
(2) MultiAttack (-Partition): This method uses Louvain's Algorithm directly for community partitioning, without using DeepWalk and K-means clustering.
(3) MultiAttack (-MultiAgent): This method does not apply the multi-agent architecture and randomly selects cold-start items and users from different communities to generate profiles. 
(4) MultiAttack (-Cold-Start): This method does not employ cold-start items to reduce the action space, and allows agents to generate item profiles by selecting from the whole items pool.

The experimental results are illustrated in Figure~\ref{fig:ablation}. Comparing MultiAttack (-Community) with MultiAttack, the improvement in attack performance indicates that incorporating cross-community connections enhances the effectiveness of the attack by creating more harmful social profiles. The comparison between MultiAttack (-Partition) and MultiAttack demonstrates that our enhanced community partitioning is more effective, improving the overall attack performance. The difference in performance between MultiAttack (-MultiAgent) and MultiAttack validates that the development of our multi-agent architecture is crucial for coordinating the social and item profile generation, resulting in a more effective attack on social recommender systems. Compared to MultiAttack (-Cold-Start), MultiAttack can significantly enhance the effectiveness of the attack. This demonstrates selecting cold-start items as fake user profiles can lead to better attack performance.
Moreover, cold-start items can effectively reduce agents' action space, therefore decreasing the learning cost.

\begin{figure*}[tb]
 \centering
\subfigure{\includegraphics[width=0.9\linewidth]{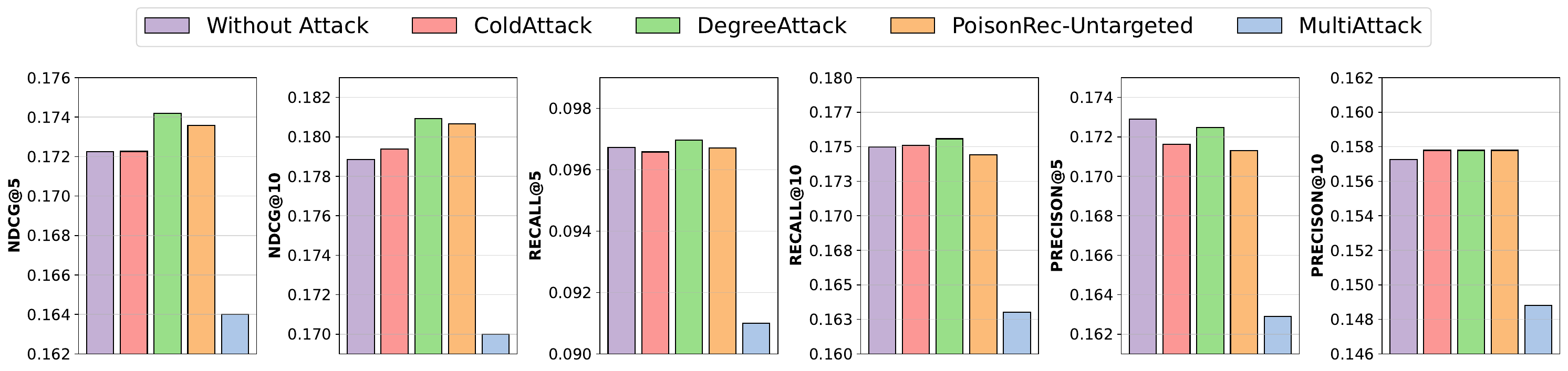}}
 \vskip -0.13in
\caption{Performance comparison of different attacking methods for inductive recommender system under defense (i.e., with adversarial training) on the LastFM dataset.}
\label{fig:APT}
\end{figure*}

\subsection{\textbf{Attack Effectiveness Under Defense}}
In this subsection, we evaluate the attack performance under defense. Specifically, we examine the attack effectiveness of MultiAttack on more robust social recommender systems (i.e., with adversarial training~\cite{he2018adversarial}). The results on the LastFM across three metrics are reported in Figure~\ref{fig:APT}. Our findings demonstrate that MultiAttack consistently outperforms the baseline model. On the other hand, vanilla attack methods cannot perform effective untargeted attacks on social recommender systems with adversarial training. In contrast, MultiAttack generates high-quality item profiles and cross-community user connections that may not be seen in adversarial training, thus effectively degrading recommendation performance.

\begin{table}[t]
  \centering
\caption{Comparison of the detected rate of fake users and recommendation performance decrease for different black-box attack methods on the LastFM dataset. Performance decrease is calculated by the metric PRECISON@5. We use bold fonts and underline to mark the lowest and second lowest detected rates and best and second best attack performance, respectively.}
    \scalebox{0.75}
{
    \begin{tabular}{c|c|c|c|c|c}
    \toprule
    \multirow{1}{*}{Attack Model}  & {RandomAttack} & {ColdAttack} & {DegreeAttack} & {PoisonRec-Untargeted} & {MultiAttack}\\
 \midrule
Detected Rate &  \textbf{4}\% & \underline{6}\% & 24\% & 22\% & \underline{6}\%\\
\midrule
 Performance Decrease& 6.3\% & 3.3\% & 6.6\% & \underline{6.9\%} &\textbf{25.6\%} \\
    \bottomrule
    \end{tabular}%
} 
  \label{tab:detection}%
\end{table}%

\subsection{\textbf{Attack Under Detection}}
In this subsection, we adopt the anomaly detection methods~\cite{zhang2019comparing,akoglu2015graph} to examine whether fake users can be easily detected. 
Specifically, we cluster users based on their social relations and interaction behaviors to detect anomalous users with low density~\cite{papadimitriou2003loci,breunig2000lof}. We compare MultiAttack with some heuristic attacks and another RL-based attack (i.e., PoisonRec-Untargeted) on the LastFM dataset. The results are shown in Table~\ref{tab:detection}.
For PoisonRec-Untargeted and DegreeAttack, 22\% and 24\% of fake users are detected.
Instead, only 6\% of fake users generated by MultiAttack are detected, which is substantially lower than PoisonRec-Untargeted.
While the detection rate of MultiAttack is slightly higher than RandomAttack, the attack effectiveness of MultiAttack substantially exceeds that of all baselines.
It suggests that our model can effectively circumvent some vanilla detection methods to conduct effective untargeted attacks for the recommender system.

\subsection{\textbf{Parameter Analysis}}
In this section, we analyze hyper-parameters used in our framework to provide insights into the effectiveness of our attacking framework.

\subsubsection{\textbf{Effect of budget}}
\begin{figure*}[t]

 \centering
 \centering
{\subfigure[NDCG@10]
{\includegraphics[width=0.32\linewidth]{{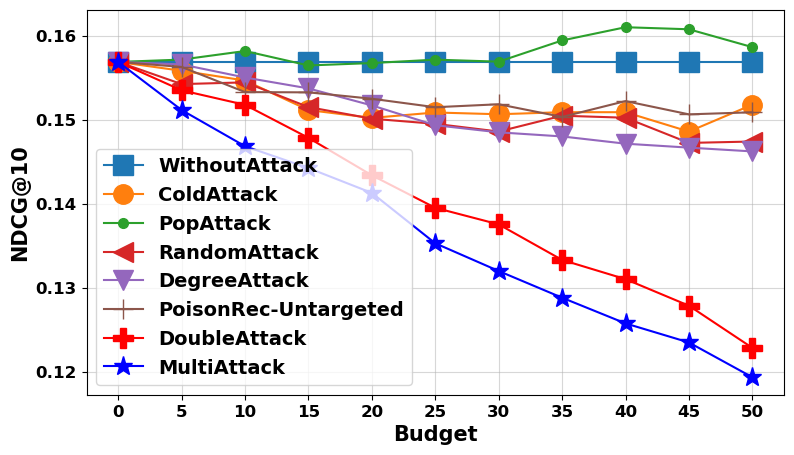}}}}
{\subfigure[RECALL@10]
{\includegraphics[width=0.32\linewidth]{{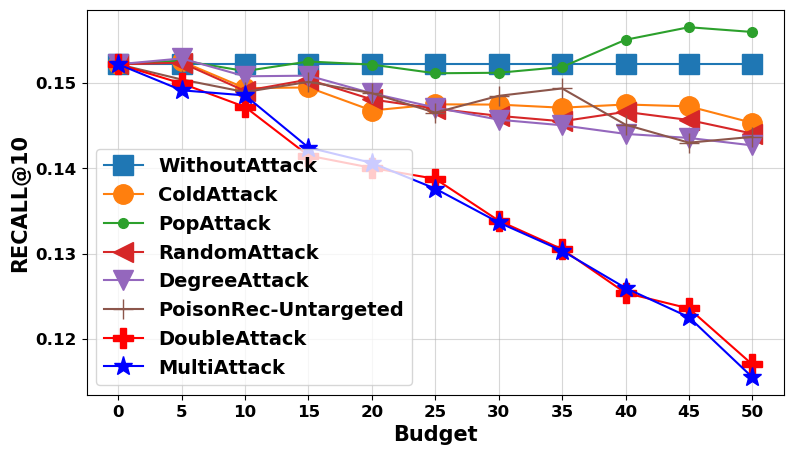}}}}
{\subfigure[PRECISION@10]
{\includegraphics[width=0.32\linewidth]{{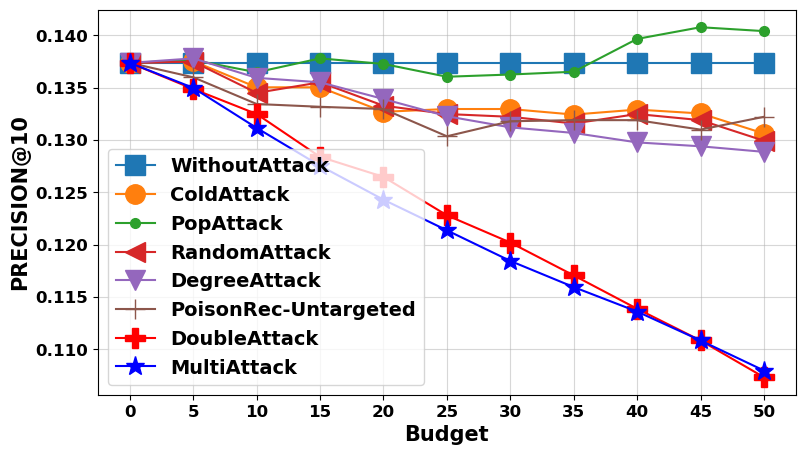}}}}
\vskip -0.12in
 \caption{Effect of budget $\Delta$~(the number of injected fake users) on LastFM across metrics NDCG@10, RECALL@10 and PRECISION@10 for inductive social recommender system (i.e., GraphSAGE with social network). }\label{fig:budget1}
 \vskip -0.060in
\end{figure*}

In this subsection, we investigate the black-box attack performance by varying the budget $\Delta$ (i.e., the number of fake users). 
We present the results for our method, MultiAttack, on the LastFM dataset across three metrics,  as shown in Figure~\ref{fig:budget1}. Overall, we can observe that our method, MultiAttack, substantially outperforms other baseline methods. 
In contrast, heuristics-based methods (e.g., RandomAttack) and PoisonRec-Untargeted hardly produce any attack improvement after injecting 25 or more fake users.
DoubleAttack shows comparable performance to MultiAttack at various budgets on Recall and Precision metrics.
On the other hand, MultiAttack can perform more effective attacks on the NDCG metric. The experimental results demonstrate that both our methods MultiAttack and DoubleAttack can effectively degrade the recommendation performance under different budgets for untargeted black-box attacks.

\subsubsection{\textbf{Effect of profile length}}
\begin{figure*}[t]
 \centering
 \vskip -0.10in
 \centering
{\subfigure[NDCG@10]
{\includegraphics[width=0.32\linewidth]{{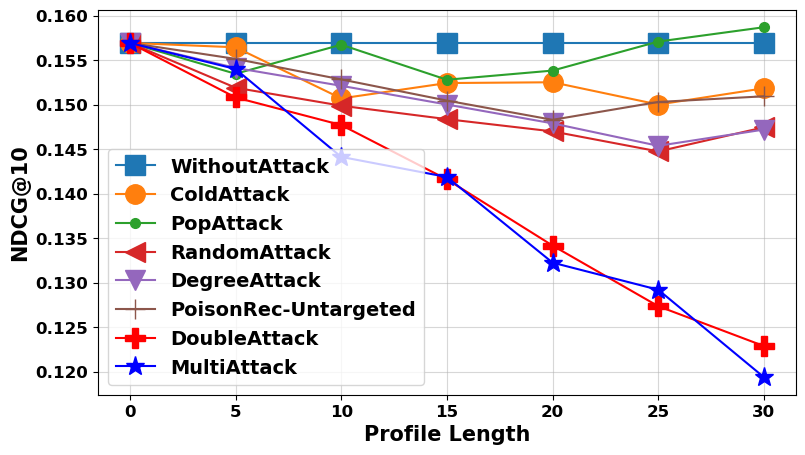}}}}
{\subfigure[RECALL@10]
{\includegraphics[width=0.32\linewidth]{{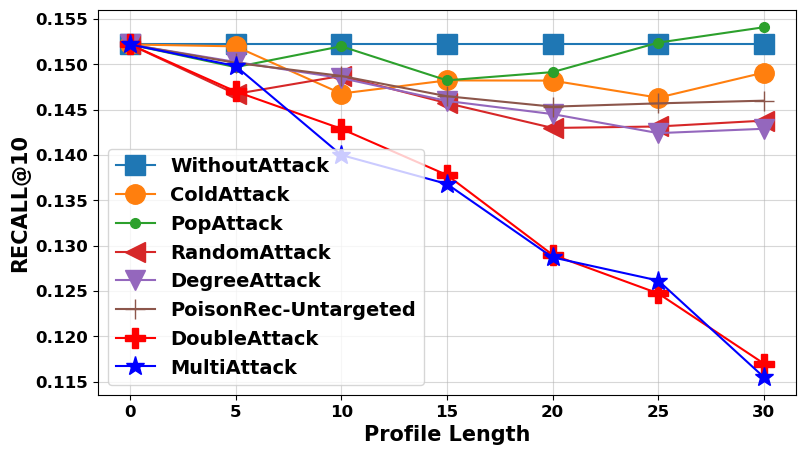}}}}
{\subfigure[PRECISION@10]
{\includegraphics[width=0.32\linewidth]{{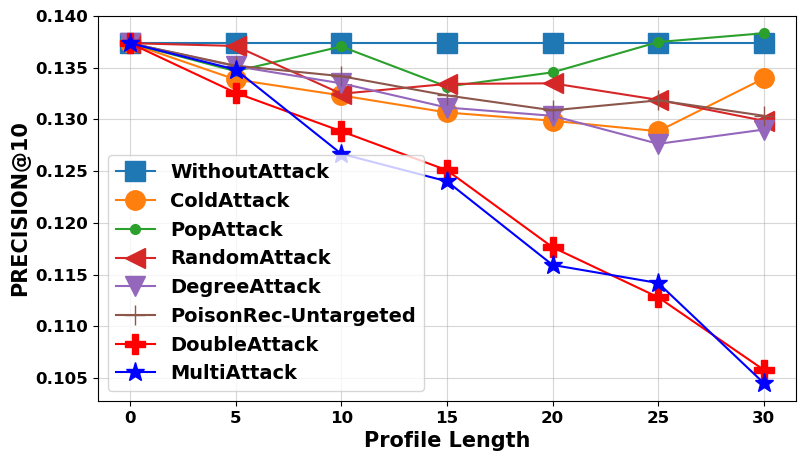}}}}
\vskip -0.12in
 \caption{Effect of profile length~(the number of clicked items) on LastFM across metrics NDCG@10, RECALL@10 and PRECISION@10 for inductive social recommender system.}\label{fig:budget2}
 \end{figure*}

In this part, we analyze the impact of profile length on the attack results and ensure that the generated user social relationship pairs remain consistent with the profile length. 
We conduct evasion attacks on GraphSAGE on the LastFM dataset using NDCG@10, Recall@10 and Precision@10 as the evaluation metrics. 
The experiment results are shown in Figure~\ref{fig:budget2}. 
In heuristic approaches, DegreeAttack considers the user's social influence, resulting in better attack performance compared to ColdAttack and PopAttack. 
In addition, DoubleAttack exhibits close performance to MultiAttack in three metrics after the profile length exceeds 15.

\begin{figure}[t]
 \centering
 \centering
{\subfigure[$k=10$]
{\includegraphics[width=0.34\linewidth]{{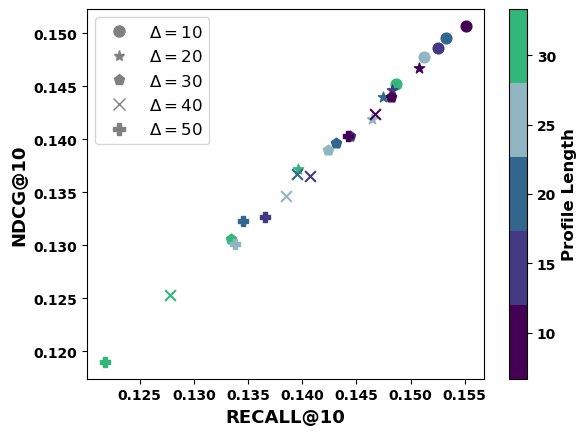}}}}
{\subfigure[$k=20$]
{\includegraphics[width=0.34\linewidth]{{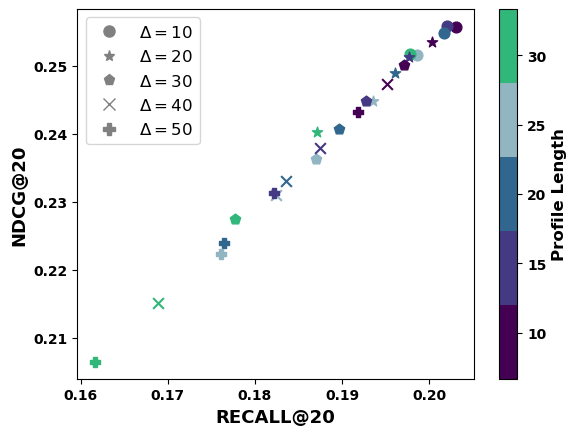}}}}
\vskip -0.12in
 \caption{The joint effect of budget $\Delta$ and profile length on LastFM across metrics NDCG@$k$ and RECALL@$k$ for inductive social recommender system.}\label{fig:budget3}
 \end{figure}

The experiment findings highlight that our decentralized policy-based approach significantly outperforms other methods when the length of the user's profile exceeds 10. 
Figure~\ref{fig:budget3} shows the joint effect of budget and item profile length on the attack.
It is crucial to select the proper profile length and budget as excessive profile length may elevate the risk of detection. Therefore, we recommend that item profiles remain within a length limit of 30.

\begin{table}[t]
  \vskip -0.08in
  \centering
  \caption{ Effect of length of epoch with NDCG@10 for inductive social recommender system.}
  \vskip -0.08in
    \scalebox{0.9}
{
    \begin{tabular}{c|c|c|c|c|c|c}
    \toprule
    \multirow{1}{*}{$K$} & 1 & 3 & 5 & 10 & 15 & 20 \\
    \midrule
LastFM & 0.12644 & 0.12434 & 0.12608 & \textbf{0.11938} & 0.12321 & 0.12570   \\
Ciao &0.11177 &0.11130 &  0.10939 & 0.10667 & \textbf{0.10612} & 0.10742 \\
Epinions &0.15632 &0.15620 &  0.15588 & 0.15525 & \textbf{0.15379} & 0.15548 \\
    \bottomrule
    \end{tabular}%
}
  \label{tab:epoch}%
    \vskip  0.05in
\end{table}%

\subsubsection{\textbf{Effect of the length of epoch}}
In this subsection, we analyze the impact of the length of epoch $K$ on the performance of our attack method, which has an impact on critic and actor networks update. More specifically, the length of the epoch determines the number of iterations for parameter updates. 
We have conducted experiments on three datasets to evaluate the optimal length of the epoch for each case. 
The experimental results are shown in Table ~\ref{tab:epoch}. 
For the small dataset, LastFM can achieve the best attacking results when $K$ sets to 10.
On the other hand, for larger datasets such as Ciao and Epinions, a setting of 15 yields the best attacking results. 
Thus, in order better train the MultiAttack, we need to carefully select epoch length for different datasets.

\subsection{\textbf{Further Analysis and Discussion}}
In this work, we propose MultiAttack to leverage multiple agents to generate cold-start item profiles and construct cross-community connections for attacking social recommender systems. Extensive experiments validate the effectiveness of the proposed method. 

It is worth noting that recommending more cold-start items might increase the \textit{exploration} rate of the recommendation model and the diversity of items. However, the trade-off between exploration and \textit{exploitation} is crucial~\cite{barraza2017exploration}. For well-trained recommendation models, exploring cold-start items extensively can lead to unnecessary costs~\cite{kadiouglu2021optimized,kadiouglu2024integrating}. In real recommendation scenarios (e.g., e-commerce platforms), over-exploration can lead to missed opportunities in competing with opponents, resulting in serious financial losses.

On the other hand, cold-start items include not only new additions but also a large number of cold-subject and restricted items. For example, an attacked film recommender system might frequently recommend documentaries to uninterested users, which can severely degrade the user experience.
Even worse, the recommender system may recommend some bloody and violent films to children. This can cause significant financial losses to the platform and even result in social harm. Therefore, it is significant and urgent to study more robust recommender systems against adversarial attacks.

%% file: sections/relatedwork.tex
\section{Related Work}
\label{sec:relatedwork}
\subsection{Social Recommendations}
With the rise of social networks, leveraging social relationships to enhance recommendation performance has attracted increasing attention in recent years. 
Early social recommendation methods such as TrustMF~\cite{yang2016social} and SoRec~\cite{ma2008sorec} utilize the matrix factorization (MF) approach to map users to two low-dimensional spaces. 
SocialMF~\cite{jamali2010matrix} improves upon this by incorporating regularization into the user-item MF to ensure that the user preference latent factor aligns with those of their social neighbors.
However, these social recommender systems only consider the influence of the user's first-order neighbors and overlook the higher-order influences in the social network.

To capture higher-order relationships in social networks, researchers develop social recommendation methods based on graph neural networks architecture.
For example, GraphRec~\cite{fan2019graph} first introduces graph neural networks to social recommendations, by modeling users' social and user-item relationships as graph data. 
DiffNet~\cite{wu2019neural} simulates how users are influenced by the recursive social diffusion process for social recommendations by leveraging GNNs for modeling. 
Recently, self-supervised contrastive learning techniques have been introduced into social recommendations to enhance the representation learning of users and items ~\cite{wu2022disentangled, du2022socially}.
DcRec~\cite{wu2022disentangled} proposes a disentangled contrastive learning framework to enhance user representations by separately learning their behavior patterns from the item and social domains.

\subsection{Adversarial Attacks on Recommendations}
Adversarial attacks on recommender systems have been studied extensively over the years ~\cite{fan2022comprehensive, fan2021attacking, song2020poisonrec}, in which attackers can manipulate recommender systems with malicious purposes by injecting adversarial (fake) user profiles. 
Some early studies ~\cite{burke2005limited, mobasher2007toward, burke2005segment} focus on attacks in a white-box setting, where fake user profiles are created and injected heuristically to influence the recommendations.
For example, bandwagon attacks ~\cite{mobasher2007toward} select popular items as filler items for fake users, and segment attacks ~\cite{burke2005segment} choose items similar to the target item as filler items to aid in the attack. 
More recent studies employ optimization-based methods ~\cite{christakopoulou2019adversarial, fang2020influence, fang2018poisoning, li2016data, tang2020revisiting} to generate fake user profiles with the intention of manipulating the recommendations, where they model the attack objective as an optimization problem and use gradient-based methods to generate fake user profiles.

However, such attacks require the attacker to possess total or partial knowledge about the recommender system, such as system's architecture\&parameters and training data, which is hard to obtain in the real world due to the protection of data privacy and security. 
Therefore, studies on black box attacks on recommender systems have recently attracted significant attention ~\cite{song2020poisonrec, chen2022knowledge, fan2021attacking, fan2023adversarial}.
PoisonRec ~\cite{song2020poisonrec} is the first black-box attacking method in recommender systems. It generates fake user profiles from the massive item sets by querying the target systems.
To explore the connectivity among items,  KGAttack~\cite{chen2022knowledge} uses the knowledge graph to enhance the generation of fake user profiles.
Instead of generating fake user profiles, CopyAttack~\cite{fan2021attacking} proposes a novel copy mechanism to obtain real user profiles from cross domains for attacking black-box recommender systems. 
Recently, FedAttack~\cite{wu2022fedattack} and ClusterAttack~\cite{yu2023untargeted} utilize clients to upload malicious gradients to servers to perform untargeted attacks on federal recommender systems.
In addition, a white-box attacking method is proposed to conduct targeted attacks on social recommender systems~\cite{hu2019targeted}. 
Despite the aforementioned success, little attention has been paid to conducting untargeted attacks on social recommendations under the black-box setting.  
To fill this gap, we propose an untargeted black-box attacking framework for social recommendations.

%% file: sections/conclusion.tex
\section{Conclusion}
\label{sec:conclusion}

In this work, we propose a novel framework \emph{MultiAttack} for conducting untargeted attacks on social recommender systems via the development of multi-agent reinforcement learning under the black-box setting. 
We first conduct several preliminary studies to demonstrate the effectiveness of cold-start items and cross-community connections in generating item profiles and social profiles for reducing the performance of recommendations. 
More specifically, MultiAttack leverages a combination of multiple agents with a centralized critic network to coordinate the generation of social profiles and item profiles for fake users. 
Extensive experiments on three real-world datasets demonstrate the superiority of our proposed MultiAttack over other baseline methods, achieving better attacking performance with different profile lengths and budgets.
Furthermore, our ablation study highlights the effectiveness of various components in the proposed method.

%% file: sample-acmsmall.bbl

\begin{thebibliography}{64}


\ifx \showCODEN    \undefined \def \showCODEN     #1{\unskip}     \fi
\ifx \showDOI      \undefined \def \showDOI       #1{#1}\fi
\ifx \showISBNx    \undefined \def \showISBNx     #1{\unskip}     \fi
\ifx \showISBNxiii \undefined \def \showISBNxiii  #1{\unskip}     \fi
\ifx \showISSN     \undefined \def \showISSN      #1{\unskip}     \fi
\ifx \showLCCN     \undefined \def \showLCCN      #1{\unskip}     \fi
\ifx \shownote     \undefined \def \shownote      #1{#1}          \fi
\ifx \showarticletitle \undefined \def \showarticletitle #1{#1}   \fi
\ifx \showURL      \undefined \def \showURL       {\relax}        \fi
\providecommand\bibfield[2]{#2}
\providecommand\bibinfo[2]{#2}
\providecommand\natexlab[1]{#1}
\providecommand\showeprint[2][]{arXiv:#2}

\bibitem[Abdollahpouri et~al\mbox{.}(2017)]%
        {abdollahpouri2017controlling}
\bibfield{author}{\bibinfo{person}{Himan Abdollahpouri}, \bibinfo{person}{Robin Burke}, {and} \bibinfo{person}{Bamshad Mobasher}.} \bibinfo{year}{2017}\natexlab{}.
\newblock \showarticletitle{Controlling popularity bias in learning-to-rank recommendation}. In \bibinfo{booktitle}{\emph{Proceedings of the eleventh ACM conference on recommender systems}}. \bibinfo{pages}{42--46}.
\newblock


\bibitem[Akoglu et~al\mbox{.}(2015)]%
        {akoglu2015graph}
\bibfield{author}{\bibinfo{person}{Leman Akoglu}, \bibinfo{person}{Hanghang Tong}, {and} \bibinfo{person}{Danai Koutra}.} \bibinfo{year}{2015}\natexlab{}.
\newblock \showarticletitle{Graph based anomaly detection and description: a survey}.
\newblock \bibinfo{journal}{\emph{Data mining and knowledge discovery}}  \bibinfo{volume}{29} (\bibinfo{year}{2015}), \bibinfo{pages}{626--688}.
\newblock


\bibitem[Barraza-Urbina(2017)]%
        {barraza2017exploration}
\bibfield{author}{\bibinfo{person}{Andrea Barraza-Urbina}.} \bibinfo{year}{2017}\natexlab{}.
\newblock \showarticletitle{The exploration-exploitation trade-off in interactive recommender systems}. In \bibinfo{booktitle}{\emph{Proceedings of the Eleventh ACM Conference on Recommender Systems}}. \bibinfo{pages}{431--435}.
\newblock


\bibitem[Blondel et~al\mbox{.}(2008)]%
        {blondel2008fast}
\bibfield{author}{\bibinfo{person}{Vincent~D Blondel}, \bibinfo{person}{Jean-Loup Guillaume}, \bibinfo{person}{Renaud Lambiotte}, {and} \bibinfo{person}{Etienne Lefebvre}.} \bibinfo{year}{2008}\natexlab{}.
\newblock \showarticletitle{Fast unfolding of communities in large networks}.
\newblock \bibinfo{journal}{\emph{Journal of statistical mechanics: theory and experiment}} \bibinfo{volume}{2008}, \bibinfo{number}{10} (\bibinfo{year}{2008}), \bibinfo{pages}{P10008}.
\newblock


\bibitem[Breunig et~al\mbox{.}(2000)]%
        {breunig2000lof}
\bibfield{author}{\bibinfo{person}{Markus~M Breunig}, \bibinfo{person}{Hans-Peter Kriegel}, \bibinfo{person}{Raymond~T Ng}, {and} \bibinfo{person}{J{\"o}rg Sander}.} \bibinfo{year}{2000}\natexlab{}.
\newblock \showarticletitle{LOF: identifying density-based local outliers}. In \bibinfo{booktitle}{\emph{Proceedings of the 2000 ACM SIGMOD international conference on Management of data}}. \bibinfo{pages}{93--104}.
\newblock


\bibitem[Brynjolfsson et~al\mbox{.}(2011)]%
        {brynjolfsson2011goodbye}
\bibfield{author}{\bibinfo{person}{Erik Brynjolfsson}, \bibinfo{person}{Yu Hu}, {and} \bibinfo{person}{Duncan Simester}.} \bibinfo{year}{2011}\natexlab{}.
\newblock \showarticletitle{Goodbye pareto principle, hello long tail: The effect of search costs on the concentration of product sales}.
\newblock \bibinfo{journal}{\emph{Management science}} \bibinfo{volume}{57}, \bibinfo{number}{8} (\bibinfo{year}{2011}), \bibinfo{pages}{1373--1386}.
\newblock


\bibitem[Burke et~al\mbox{.}(2005a)]%
        {burke2005limited}
\bibfield{author}{\bibinfo{person}{Robin Burke}, \bibinfo{person}{Bamshad Mobasher}, {and} \bibinfo{person}{Runa Bhaumik}.} \bibinfo{year}{2005}\natexlab{a}.
\newblock \showarticletitle{Limited knowledge shilling attacks in collaborative filtering systems}. In \bibinfo{booktitle}{\emph{Proceedings of 3rd international workshop on intelligent techniques for web personalization (ITWP 2005), 19th international joint conference on artificial intelligence (IJCAI 2005)}}. \bibinfo{pages}{17--24}.
\newblock


\bibitem[Burke et~al\mbox{.}(2005b)]%
        {burke2005segment}
\bibfield{author}{\bibinfo{person}{Robin Burke}, \bibinfo{person}{Bamshad Mobasher}, \bibinfo{person}{Runa Bhaumik}, {and} \bibinfo{person}{Chad Williams}.} \bibinfo{year}{2005}\natexlab{b}.
\newblock \showarticletitle{Segment-based injection attacks against collaborative filtering recommender systems}. In \bibinfo{booktitle}{\emph{Fifth IEEE International Conference on Data Mining (ICDM'05)}}. IEEE, \bibinfo{pages}{4--pp}.
\newblock


\bibitem[Bu{\c{s}}oniu et~al\mbox{.}(2010)]%
        {bucsoniu2010multi}
\bibfield{author}{\bibinfo{person}{Lucian Bu{\c{s}}oniu}, \bibinfo{person}{Robert Babu{\v{s}}ka}, {and} \bibinfo{person}{Bart De~Schutter}.} \bibinfo{year}{2010}\natexlab{}.
\newblock \showarticletitle{Multi-agent reinforcement learning: An overview}.
\newblock \bibinfo{journal}{\emph{Innovations in multi-agent systems and applications-1}} (\bibinfo{year}{2010}), \bibinfo{pages}{183--221}.
\newblock


\bibitem[Celma and Herrera(2008)]%
        {celma2008new}
\bibfield{author}{\bibinfo{person}{{\`O}scar Celma} {and} \bibinfo{person}{Perfecto Herrera}.} \bibinfo{year}{2008}\natexlab{}.
\newblock \showarticletitle{A new approach to evaluating novel recommendations}. In \bibinfo{booktitle}{\emph{Proceedings of the 2008 ACM conference on Recommender systems}}. \bibinfo{pages}{179--186}.
\newblock


\bibitem[Chen et~al\mbox{.}(2022)]%
        {chen2022knowledge}
\bibfield{author}{\bibinfo{person}{Jingfan Chen}, \bibinfo{person}{Wenqi Fan}, \bibinfo{person}{Guanghui Zhu}, \bibinfo{person}{Xiangyu Zhao}, \bibinfo{person}{Chunfeng Yuan}, \bibinfo{person}{Qing Li}, {and} \bibinfo{person}{Yihua Huang}.} \bibinfo{year}{2022}\natexlab{}.
\newblock \showarticletitle{Knowledge-enhanced Black-box Attacks for Recommendations}. In \bibinfo{booktitle}{\emph{Proceedings of the 28th ACM SIGKDD Conference on Knowledge Discovery and Data Mining}}. \bibinfo{pages}{108--117}.
\newblock


\bibitem[Cheng et~al\mbox{.}(2016)]%
        {cheng2016wide}
\bibfield{author}{\bibinfo{person}{Heng-Tze Cheng}, \bibinfo{person}{Levent Koc}, \bibinfo{person}{Jeremiah Harmsen}, \bibinfo{person}{Tal Shaked}, \bibinfo{person}{Tushar Chandra}, \bibinfo{person}{Hrishi Aradhye}, \bibinfo{person}{Glen Anderson}, \bibinfo{person}{Greg Corrado}, \bibinfo{person}{Wei Chai}, \bibinfo{person}{Mustafa Ispir}, {et~al\mbox{.}}} \bibinfo{year}{2016}\natexlab{}.
\newblock \showarticletitle{Wide \& deep learning for recommender systems}. In \bibinfo{booktitle}{\emph{Proceedings of the 1st workshop on deep learning for recommender systems}}. \bibinfo{pages}{7--10}.
\newblock


\bibitem[Christakopoulou and Banerjee(2019)]%
        {christakopoulou2019adversarial}
\bibfield{author}{\bibinfo{person}{Konstantina Christakopoulou} {and} \bibinfo{person}{Arindam Banerjee}.} \bibinfo{year}{2019}\natexlab{}.
\newblock \showarticletitle{Adversarial attacks on an oblivious recommender}. In \bibinfo{booktitle}{\emph{Proceedings of the 13th ACM Conference on Recommender Systems}}. \bibinfo{pages}{322--330}.
\newblock


\bibitem[Du et~al\mbox{.}(2022)]%
        {du2022socially}
\bibfield{author}{\bibinfo{person}{Jing Du}, \bibinfo{person}{Zesheng Ye}, \bibinfo{person}{Lina Yao}, \bibinfo{person}{Bin Guo}, {and} \bibinfo{person}{Zhiwen Yu}.} \bibinfo{year}{2022}\natexlab{}.
\newblock \showarticletitle{Socially-aware dual contrastive learning for cold-start recommendation}. In \bibinfo{booktitle}{\emph{Proceedings of the 45th International ACM SIGIR Conference on Research and Development in Information Retrieval}}. \bibinfo{pages}{1927--1932}.
\newblock


\bibitem[Fan et~al\mbox{.}(2021)]%
        {fan2021attacking}
\bibfield{author}{\bibinfo{person}{Wenqi Fan}, \bibinfo{person}{Tyler Derr}, \bibinfo{person}{Xiangyu Zhao}, \bibinfo{person}{Yao Ma}, \bibinfo{person}{Hui Liu}, \bibinfo{person}{Jianping Wang}, \bibinfo{person}{Jiliang Tang}, {and} \bibinfo{person}{Qing Li}.} \bibinfo{year}{2021}\natexlab{}.
\newblock \showarticletitle{Attacking black-box recommendations via copying cross-domain user profiles}. In \bibinfo{booktitle}{\emph{2021 IEEE 37th International Conference on Data Engineering (ICDE)}}. IEEE, \bibinfo{pages}{1583--1594}.
\newblock


\bibitem[Fan et~al\mbox{.}(2018)]%
        {fan2018deep}
\bibfield{author}{\bibinfo{person}{Wenqi Fan}, \bibinfo{person}{Qing Li}, {and} \bibinfo{person}{Min Cheng}.} \bibinfo{year}{2018}\natexlab{}.
\newblock \showarticletitle{Deep modeling of social relations for recommendation}. In \bibinfo{booktitle}{\emph{Proceedings of the AAAI Conference on Artificial Intelligence}}, Vol.~\bibinfo{volume}{32}.
\newblock


\bibitem[Fan et~al\mbox{.}(2019a)]%
        {fan2019graph}
\bibfield{author}{\bibinfo{person}{Wenqi Fan}, \bibinfo{person}{Yao Ma}, \bibinfo{person}{Qing Li}, \bibinfo{person}{Yuan He}, \bibinfo{person}{Eric Zhao}, \bibinfo{person}{Jiliang Tang}, {and} \bibinfo{person}{Dawei Yin}.} \bibinfo{year}{2019}\natexlab{a}.
\newblock \showarticletitle{Graph neural networks for social recommendation}. In \bibinfo{booktitle}{\emph{The world wide web conference}}. \bibinfo{pages}{417--426}.
\newblock


\bibitem[Fan et~al\mbox{.}(2019b)]%
        {fan2019deep_dscf}
\bibfield{author}{\bibinfo{person}{Wenqi Fan}, \bibinfo{person}{Yao Ma}, \bibinfo{person}{Dawei Yin}, \bibinfo{person}{Jianping Wang}, \bibinfo{person}{Jiliang Tang}, {and} \bibinfo{person}{Qing Li}.} \bibinfo{year}{2019}\natexlab{b}.
\newblock \showarticletitle{Deep social collaborative filtering}. In \bibinfo{booktitle}{\emph{Proceedings of the 13th ACM conference on recommender systems}}. \bibinfo{pages}{305--313}.
\newblock


\bibitem[Fan et~al\mbox{.}(2022)]%
        {fan2022comprehensive}
\bibfield{author}{\bibinfo{person}{Wenqi Fan}, \bibinfo{person}{Xiangyu Zhao}, \bibinfo{person}{Xiao Chen}, \bibinfo{person}{Jingran Su}, \bibinfo{person}{Jingtong Gao}, \bibinfo{person}{Lin Wang}, \bibinfo{person}{Qidong Liu}, \bibinfo{person}{Yiqi Wang}, \bibinfo{person}{Han Xu}, \bibinfo{person}{Lei Chen}, {et~al\mbox{.}}} \bibinfo{year}{2022}\natexlab{}.
\newblock \showarticletitle{A Comprehensive Survey on Trustworthy Recommender Systems}.
\newblock \bibinfo{journal}{\emph{arXiv preprint arXiv:2209.10117}} (\bibinfo{year}{2022}).
\newblock


\bibitem[Fan et~al\mbox{.}(2023)]%
        {fan2023adversarial}
\bibfield{author}{\bibinfo{person}{Wenqi Fan}, \bibinfo{person}{Xiangyu Zhao}, \bibinfo{person}{Qing Li}, \bibinfo{person}{Tyler Derr}, \bibinfo{person}{Yao Ma}, \bibinfo{person}{Hui Liu}, \bibinfo{person}{Jianping Wang}, {and} \bibinfo{person}{Jiliang Tang}.} \bibinfo{year}{2023}\natexlab{}.
\newblock \showarticletitle{Adversarial Attacks for Black-Box Recommender Systems Via Copying Transferable Cross-Domain User Profiles}.
\newblock \bibinfo{journal}{\emph{IEEE Transactions on Knowledge and Data Engineering}} (\bibinfo{year}{2023}).
\newblock


\bibitem[Fang et~al\mbox{.}(2020)]%
        {fang2020influence}
\bibfield{author}{\bibinfo{person}{Minghong Fang}, \bibinfo{person}{Neil~Zhenqiang Gong}, {and} \bibinfo{person}{Jia Liu}.} \bibinfo{year}{2020}\natexlab{}.
\newblock \showarticletitle{Influence function based data poisoning attacks to top-n recommender systems}. In \bibinfo{booktitle}{\emph{Proceedings of The Web Conference 2020}}. \bibinfo{pages}{3019--3025}.
\newblock


\bibitem[Fang et~al\mbox{.}(2018)]%
        {fang2018poisoning}
\bibfield{author}{\bibinfo{person}{Minghong Fang}, \bibinfo{person}{Guolei Yang}, \bibinfo{person}{Neil~Zhenqiang Gong}, {and} \bibinfo{person}{Jia Liu}.} \bibinfo{year}{2018}\natexlab{}.
\newblock \showarticletitle{Poisoning attacks to graph-based recommender systems}. In \bibinfo{booktitle}{\emph{Proceedings of the 34th annual computer security applications conference}}. \bibinfo{pages}{381--392}.
\newblock


\bibitem[Gasparetti et~al\mbox{.}(2021)]%
        {gasparetti2021community}
\bibfield{author}{\bibinfo{person}{Fabio Gasparetti}, \bibinfo{person}{Giuseppe Sansonetti}, {and} \bibinfo{person}{Alessandro Micarelli}.} \bibinfo{year}{2021}\natexlab{}.
\newblock \showarticletitle{Community detection in social recommender systems: a survey}.
\newblock \bibinfo{journal}{\emph{Applied Intelligence}}  \bibinfo{volume}{51} (\bibinfo{year}{2021}), \bibinfo{pages}{3975--3995}.
\newblock


\bibitem[Gupta et~al\mbox{.}(2017)]%
        {gupta2017cooperative}
\bibfield{author}{\bibinfo{person}{Jayesh~K Gupta}, \bibinfo{person}{Maxim Egorov}, {and} \bibinfo{person}{Mykel Kochenderfer}.} \bibinfo{year}{2017}\natexlab{}.
\newblock \showarticletitle{Cooperative multi-agent control using deep reinforcement learning}. In \bibinfo{booktitle}{\emph{Autonomous Agents and Multiagent Systems: AAMAS 2017 Workshops, Best Papers, S{\~a}o Paulo, Brazil, May 8-12, 2017, Revised Selected Papers 16}}. Springer, \bibinfo{pages}{66--83}.
\newblock


\bibitem[Hamilton et~al\mbox{.}(2017)]%
        {hamilton2017inductive}
\bibfield{author}{\bibinfo{person}{Will Hamilton}, \bibinfo{person}{Zhitao Ying}, {and} \bibinfo{person}{Jure Leskovec}.} \bibinfo{year}{2017}\natexlab{}.
\newblock \showarticletitle{Inductive representation learning on large graphs}.
\newblock \bibinfo{journal}{\emph{Advances in neural information processing systems}}  \bibinfo{volume}{30} (\bibinfo{year}{2017}).
\newblock


\bibitem[He et~al\mbox{.}(2020)]%
        {He2020LightGCNSA}
\bibfield{author}{\bibinfo{person}{Xiangnan He}, \bibinfo{person}{Kuan Deng}, \bibinfo{person}{X. Wang}, \bibinfo{person}{Y. Li}, \bibinfo{person}{Yongdong Zhang}, {and} \bibinfo{person}{Meng Wang}.} \bibinfo{year}{2020}\natexlab{}.
\newblock \showarticletitle{LightGCN: Simplifying and Powering Graph Convolution Network for Recommendation}.
\newblock \bibinfo{journal}{\emph{ACM SIGIR}} (\bibinfo{year}{2020}).
\newblock


\bibitem[He et~al\mbox{.}(2018)]%
        {he2018adversarial}
\bibfield{author}{\bibinfo{person}{Xiangnan He}, \bibinfo{person}{Zhankui He}, \bibinfo{person}{Xiaoyu Du}, {and} \bibinfo{person}{Tat-Seng Chua}.} \bibinfo{year}{2018}\natexlab{}.
\newblock \showarticletitle{Adversarial personalized ranking for recommendation}. In \bibinfo{booktitle}{\emph{The 41st International ACM SIGIR conference on research \& development in information retrieval}}. \bibinfo{pages}{355--364}.
\newblock


\bibitem[Hessel et~al\mbox{.}(2019)]%
        {hessel2019multi}
\bibfield{author}{\bibinfo{person}{Matteo Hessel}, \bibinfo{person}{Hubert Soyer}, \bibinfo{person}{Lasse Espeholt}, \bibinfo{person}{Wojciech Czarnecki}, \bibinfo{person}{Simon Schmitt}, {and} \bibinfo{person}{Hado Van~Hasselt}.} \bibinfo{year}{2019}\natexlab{}.
\newblock \showarticletitle{Multi-task deep reinforcement learning with popart}. In \bibinfo{booktitle}{\emph{Proceedings of the AAAI Conference on Artificial Intelligence}}, Vol.~\bibinfo{volume}{33}. \bibinfo{pages}{3796--3803}.
\newblock


\bibitem[Hu et~al\mbox{.}(2019)]%
        {hu2019targeted}
\bibfield{author}{\bibinfo{person}{Rui Hu}, \bibinfo{person}{Yuanxiong Guo}, \bibinfo{person}{Miao Pan}, {and} \bibinfo{person}{Yanmin Gong}.} \bibinfo{year}{2019}\natexlab{}.
\newblock \showarticletitle{Targeted poisoning attacks on social recommender systems}. In \bibinfo{booktitle}{\emph{2019 IEEE Global Communications Conference (GLOBECOM)}}. IEEE, \bibinfo{pages}{1--6}.
\newblock


\bibitem[Jamali and Ester(2010)]%
        {jamali2010matrix}
\bibfield{author}{\bibinfo{person}{Mohsen Jamali} {and} \bibinfo{person}{Martin Ester}.} \bibinfo{year}{2010}\natexlab{}.
\newblock \showarticletitle{A matrix factorization technique with trust propagation for recommendation in social networks}. In \bibinfo{booktitle}{\emph{Proceedings of the fourth ACM conference on Recommender systems}}. \bibinfo{pages}{135--142}.
\newblock


\bibitem[Kad{\i}o{\u{g}}lu et~al\mbox{.}(2021)]%
        {kadiouglu2021optimized}
\bibfield{author}{\bibinfo{person}{Serdar Kad{\i}o{\u{g}}lu}, \bibinfo{person}{Bernard Kleynhans}, {and} \bibinfo{person}{Xin Wang}.} \bibinfo{year}{2021}\natexlab{}.
\newblock \showarticletitle{Optimized item selection to boost exploration for recommender systems}. In \bibinfo{booktitle}{\emph{Integration of Constraint Programming, Artificial Intelligence, and Operations Research: 18th International Conference, CPAIOR 2021, Vienna, Austria, July 5--8, 2021, Proceedings 18}}. Springer, \bibinfo{pages}{427--445}.
\newblock


\bibitem[Kad{\i}o{\u{g}}lu et~al\mbox{.}(2024)]%
        {kadiouglu2024integrating}
\bibfield{author}{\bibinfo{person}{Serdar Kad{\i}o{\u{g}}lu}, \bibinfo{person}{Bernard Kleynhans}, {and} \bibinfo{person}{Xin Wang}.} \bibinfo{year}{2024}\natexlab{}.
\newblock \showarticletitle{Integrating optimized item selection with active learning for continuous exploration in recommender systems}.
\newblock \bibinfo{journal}{\emph{Annals of Mathematics and Artificial Intelligence}} (\bibinfo{year}{2024}), \bibinfo{pages}{1--23}.
\newblock


\bibitem[Li et~al\mbox{.}(2016)]%
        {li2016data}
\bibfield{author}{\bibinfo{person}{Bo Li}, \bibinfo{person}{Yining Wang}, \bibinfo{person}{Aarti Singh}, {and} \bibinfo{person}{Yevgeniy Vorobeychik}.} \bibinfo{year}{2016}\natexlab{}.
\newblock \showarticletitle{Data poisoning attacks on factorization-based collaborative filtering}.
\newblock \bibinfo{journal}{\emph{Advances in neural information processing systems}}  \bibinfo{volume}{29} (\bibinfo{year}{2016}).
\newblock


\bibitem[Liao et~al\mbox{.}(2022)]%
        {liao2022sociallgn}
\bibfield{author}{\bibinfo{person}{Jie Liao}, \bibinfo{person}{Wei Zhou}, \bibinfo{person}{Fengji Luo}, \bibinfo{person}{Junhao Wen}, \bibinfo{person}{Min Gao}, \bibinfo{person}{Xiuhua Li}, {and} \bibinfo{person}{Jun Zeng}.} \bibinfo{year}{2022}\natexlab{}.
\newblock \showarticletitle{SocialLGN: Light graph convolution network for social recommendation}.
\newblock \bibinfo{journal}{\emph{Information Sciences}}  \bibinfo{volume}{589} (\bibinfo{year}{2022}), \bibinfo{pages}{595--607}.
\newblock


\bibitem[Ma et~al\mbox{.}(2009)]%
        {ma2009learning}
\bibfield{author}{\bibinfo{person}{Hao Ma}, \bibinfo{person}{Irwin King}, {and} \bibinfo{person}{Michael~R Lyu}.} \bibinfo{year}{2009}\natexlab{}.
\newblock \showarticletitle{Learning to recommend with social trust ensemble}. In \bibinfo{booktitle}{\emph{Proceedings of the 32nd international ACM SIGIR conference on Research and development in information retrieval}}. \bibinfo{pages}{203--210}.
\newblock


\bibitem[Ma et~al\mbox{.}(2008)]%
        {ma2008sorec}
\bibfield{author}{\bibinfo{person}{Hao Ma}, \bibinfo{person}{Haixuan Yang}, \bibinfo{person}{Michael~R Lyu}, {and} \bibinfo{person}{Irwin King}.} \bibinfo{year}{2008}\natexlab{}.
\newblock \showarticletitle{Sorec: social recommendation using probabilistic matrix factorization}. In \bibinfo{booktitle}{\emph{Proceedings of the 17th ACM conference on Information and knowledge management}}. \bibinfo{pages}{931--940}.
\newblock


\bibitem[Mnih et~al\mbox{.}(2016)]%
        {mnih2016asynchronous}
\bibfield{author}{\bibinfo{person}{Volodymyr Mnih}, \bibinfo{person}{Adria~Puigdomenech Badia}, \bibinfo{person}{Mehdi Mirza}, \bibinfo{person}{Alex Graves}, \bibinfo{person}{Timothy Lillicrap}, \bibinfo{person}{Tim Harley}, \bibinfo{person}{David Silver}, {and} \bibinfo{person}{Koray Kavukcuoglu}.} \bibinfo{year}{2016}\natexlab{}.
\newblock \showarticletitle{Asynchronous methods for deep reinforcement learning}. In \bibinfo{booktitle}{\emph{International conference on machine learning}}. PMLR, \bibinfo{pages}{1928--1937}.
\newblock


\bibitem[Mobasher et~al\mbox{.}(2007)]%
        {mobasher2007toward}
\bibfield{author}{\bibinfo{person}{Bamshad Mobasher}, \bibinfo{person}{Robin Burke}, \bibinfo{person}{Runa Bhaumik}, {and} \bibinfo{person}{Chad Williams}.} \bibinfo{year}{2007}\natexlab{}.
\newblock \showarticletitle{Toward trustworthy recommender systems: An analysis of attack models and algorithm robustness}.
\newblock \bibinfo{journal}{\emph{ACM Transactions on Internet Technology (TOIT)}} \bibinfo{volume}{7}, \bibinfo{number}{4} (\bibinfo{year}{2007}), \bibinfo{pages}{23--es}.
\newblock


\bibitem[Papadimitriou et~al\mbox{.}(2003)]%
        {papadimitriou2003loci}
\bibfield{author}{\bibinfo{person}{Spiros Papadimitriou}, \bibinfo{person}{Hiroyuki Kitagawa}, \bibinfo{person}{Phillip~B Gibbons}, {and} \bibinfo{person}{Christos Faloutsos}.} \bibinfo{year}{2003}\natexlab{}.
\newblock \showarticletitle{Loci: Fast outlier detection using the local correlation integral}. In \bibinfo{booktitle}{\emph{Proceedings 19th international conference on data engineering (Cat. No. 03CH37405)}}. IEEE, \bibinfo{pages}{315--326}.
\newblock


\bibitem[Perozzi et~al\mbox{.}(2014)]%
        {perozzi2014deepwalk}
\bibfield{author}{\bibinfo{person}{Bryan Perozzi}, \bibinfo{person}{Rami Al-Rfou}, {and} \bibinfo{person}{Steven Skiena}.} \bibinfo{year}{2014}\natexlab{}.
\newblock \showarticletitle{Deepwalk: Online learning of social representations}. In \bibinfo{booktitle}{\emph{Proceedings of the 20th ACM SIGKDD international conference on Knowledge discovery and data mining}}. \bibinfo{pages}{701--710}.
\newblock


\bibitem[Qu et~al\mbox{.}(2024)]%
        {qu2024tokenrec}
\bibfield{author}{\bibinfo{person}{Haohao Qu}, \bibinfo{person}{Wenqi Fan}, \bibinfo{person}{Zihuai Zhao}, {and} \bibinfo{person}{Qing Li}.} \bibinfo{year}{2024}\natexlab{}.
\newblock \showarticletitle{TokenRec: Learning to Tokenize ID for LLM-based Generative Recommendation}.
\newblock \bibinfo{journal}{\emph{arXiv preprint arXiv:2406.10450}} (\bibinfo{year}{2024}).
\newblock


\bibitem[Rendle et~al\mbox{.}(2012)]%
        {rendle2012bpr}
\bibfield{author}{\bibinfo{person}{Steffen Rendle}, \bibinfo{person}{Christoph Freudenthaler}, \bibinfo{person}{Zeno Gantner}, {and} \bibinfo{person}{Lars Schmidt-Thieme}.} \bibinfo{year}{2012}\natexlab{}.
\newblock \showarticletitle{BPR: Bayesian personalized ranking from implicit feedback}.
\newblock \bibinfo{journal}{\emph{arXiv preprint arXiv:1205.2618}} (\bibinfo{year}{2012}).
\newblock


\bibitem[Schulman et~al\mbox{.}(2017)]%
        {schulman2017proximal}
\bibfield{author}{\bibinfo{person}{John Schulman}, \bibinfo{person}{Filip Wolski}, \bibinfo{person}{Prafulla Dhariwal}, \bibinfo{person}{Alec Radford}, {and} \bibinfo{person}{Oleg Klimov}.} \bibinfo{year}{2017}\natexlab{}.
\newblock \showarticletitle{Proximal policy optimization algorithms}.
\newblock \bibinfo{journal}{\emph{arXiv preprint arXiv:1707.06347}} (\bibinfo{year}{2017}).
\newblock


\bibitem[Song et~al\mbox{.}(2020)]%
        {song2020poisonrec}
\bibfield{author}{\bibinfo{person}{Junshuai Song}, \bibinfo{person}{Zhao Li}, \bibinfo{person}{Zehong Hu}, \bibinfo{person}{Yucheng Wu}, \bibinfo{person}{Zhenpeng Li}, \bibinfo{person}{Jian Li}, {and} \bibinfo{person}{Jun Gao}.} \bibinfo{year}{2020}\natexlab{}.
\newblock \showarticletitle{Poisonrec: an adaptive data poisoning framework for attacking black-box recommender systems}. In \bibinfo{booktitle}{\emph{2020 IEEE 36th International Conference on Data Engineering (ICDE)}}. IEEE, \bibinfo{pages}{157--168}.
\newblock


\bibitem[Tang et~al\mbox{.}(2016)]%
        {tang2016recommendation}
\bibfield{author}{\bibinfo{person}{Jiliang Tang}, \bibinfo{person}{Suhang Wang}, \bibinfo{person}{Xia Hu}, \bibinfo{person}{Dawei Yin}, \bibinfo{person}{Yingzhou Bi}, \bibinfo{person}{Yi Chang}, {and} \bibinfo{person}{Huan Liu}.} \bibinfo{year}{2016}\natexlab{}.
\newblock \showarticletitle{Recommendation with social dimensions}. In \bibinfo{booktitle}{\emph{Proceedings of the AAAI conference on artificial intelligence}}, Vol.~\bibinfo{volume}{30}.
\newblock


\bibitem[Tang et~al\mbox{.}(2020)]%
        {tang2020revisiting}
\bibfield{author}{\bibinfo{person}{Jiaxi Tang}, \bibinfo{person}{Hongyi Wen}, {and} \bibinfo{person}{Ke Wang}.} \bibinfo{year}{2020}\natexlab{}.
\newblock \showarticletitle{Revisiting adversarially learned injection attacks against recommender systems}. In \bibinfo{booktitle}{\emph{Proceedings of the 14th ACM Conference on Recommender Systems}}. \bibinfo{pages}{318--327}.
\newblock


\bibitem[Tesauro(1991)]%
        {tesauro1991practical}
\bibfield{author}{\bibinfo{person}{Gerald Tesauro}.} \bibinfo{year}{1991}\natexlab{}.
\newblock \showarticletitle{Practical issues in temporal difference learning}.
\newblock \bibinfo{journal}{\emph{Advances in neural information processing systems}}  \bibinfo{volume}{4} (\bibinfo{year}{1991}).
\newblock


\bibitem[Tesauro et~al\mbox{.}(1995)]%
        {tesauro1995temporal}
\bibfield{author}{\bibinfo{person}{Gerald Tesauro} {et~al\mbox{.}}} \bibinfo{year}{1995}\natexlab{}.
\newblock \showarticletitle{Temporal difference learning and TD-Gammon}.
\newblock \bibinfo{journal}{\emph{Commun. ACM}} \bibinfo{volume}{38}, \bibinfo{number}{3} (\bibinfo{year}{1995}), \bibinfo{pages}{58--68}.
\newblock


\bibitem[Traag et~al\mbox{.}(2019)]%
        {traag2019louvain}
\bibfield{author}{\bibinfo{person}{Vincent~A Traag}, \bibinfo{person}{Ludo Waltman}, {and} \bibinfo{person}{Nees~Jan Van~Eck}.} \bibinfo{year}{2019}\natexlab{}.
\newblock \showarticletitle{From Louvain to Leiden: guaranteeing well-connected communities}.
\newblock \bibinfo{journal}{\emph{Scientific reports}} \bibinfo{volume}{9}, \bibinfo{number}{1} (\bibinfo{year}{2019}), \bibinfo{pages}{5233}.
\newblock


\bibitem[Wang and Wang(2023)]%
        {wang2023novel}
\bibfield{author}{\bibinfo{person}{Shijie Wang} {and} \bibinfo{person}{Shangbo Wang}.} \bibinfo{year}{2023}\natexlab{}.
\newblock \showarticletitle{A Novel Multi-Agent Deep RL Approach for Traffic Signal Control}. In \bibinfo{booktitle}{\emph{2023 IEEE International Conference on Pervasive Computing and Communications Workshops and other Affiliated Events (PerCom Workshops)}}. IEEE, \bibinfo{pages}{15--20}.
\newblock


\bibitem[Wang et~al\mbox{.}(2019)]%
        {Wang2019NeuralGC}
\bibfield{author}{\bibinfo{person}{Xiang Wang}, \bibinfo{person}{Xiangnan He}, \bibinfo{person}{Meng Wang}, \bibinfo{person}{Fuli Feng}, {and} \bibinfo{person}{Tat-Seng Chua}.} \bibinfo{year}{2019}\natexlab{}.
\newblock \showarticletitle{Neural Graph Collaborative Filtering}.
\newblock \bibinfo{journal}{\emph{ACM SIGIR}} (\bibinfo{year}{2019}).
\newblock


\bibitem[Wu et~al\mbox{.}(2022b)]%
        {wu2022fedattack}
\bibfield{author}{\bibinfo{person}{Chuhan Wu}, \bibinfo{person}{Fangzhao Wu}, \bibinfo{person}{Tao Qi}, \bibinfo{person}{Yongfeng Huang}, {and} \bibinfo{person}{Xing Xie}.} \bibinfo{year}{2022}\natexlab{b}.
\newblock \showarticletitle{FedAttack: Effective and covert poisoning attack on federated recommendation via hard sampling}. In \bibinfo{booktitle}{\emph{Proceedings of the 28th ACM SIGKDD Conference on Knowledge Discovery and Data Mining}}. \bibinfo{pages}{4164--4172}.
\newblock


\bibitem[Wu et~al\mbox{.}(2022a)]%
        {wu2022disentangled}
\bibfield{author}{\bibinfo{person}{Jiahao Wu}, \bibinfo{person}{Wenqi Fan}, \bibinfo{person}{Jingfan Chen}, \bibinfo{person}{Shengcai Liu}, \bibinfo{person}{Qing Li}, {and} \bibinfo{person}{Ke Tang}.} \bibinfo{year}{2022}\natexlab{a}.
\newblock \showarticletitle{Disentangled contrastive learning for social recommendation}. In \bibinfo{booktitle}{\emph{Proceedings of the 31st ACM International Conference on Information \& Knowledge Management}}. \bibinfo{pages}{4570--4574}.
\newblock


\bibitem[Wu et~al\mbox{.}(2019)]%
        {wu2019neural}
\bibfield{author}{\bibinfo{person}{Le Wu}, \bibinfo{person}{Peijie Sun}, \bibinfo{person}{Yanjie Fu}, \bibinfo{person}{Richang Hong}, \bibinfo{person}{Xiting Wang}, {and} \bibinfo{person}{Meng Wang}.} \bibinfo{year}{2019}\natexlab{}.
\newblock \showarticletitle{A neural influence diffusion model for social recommendation}. In \bibinfo{booktitle}{\emph{Proceedings of the 42nd international ACM SIGIR conference on research and development in information retrieval}}. \bibinfo{pages}{235--244}.
\newblock


\bibitem[Wu et~al\mbox{.}(2021)]%
        {wu2021coordinated}
\bibfield{author}{\bibinfo{person}{Zifan Wu}, \bibinfo{person}{Chao Yu}, \bibinfo{person}{Deheng Ye}, \bibinfo{person}{Junge Zhang}, \bibinfo{person}{Hankz~Hankui Zhuo}, {et~al\mbox{.}}} \bibinfo{year}{2021}\natexlab{}.
\newblock \showarticletitle{Coordinated proximal policy optimization}.
\newblock \bibinfo{journal}{\emph{Advances in Neural Information Processing Systems}}  \bibinfo{volume}{34} (\bibinfo{year}{2021}), \bibinfo{pages}{26437--26448}.
\newblock


\bibitem[Yang et~al\mbox{.}(2016)]%
        {yang2016social}
\bibfield{author}{\bibinfo{person}{Bo Yang}, \bibinfo{person}{Yu Lei}, \bibinfo{person}{Jiming Liu}, {and} \bibinfo{person}{Wenjie Li}.} \bibinfo{year}{2016}\natexlab{}.
\newblock \showarticletitle{Social collaborative filtering by trust}.
\newblock \bibinfo{journal}{\emph{IEEE transactions on pattern analysis and machine intelligence}} \bibinfo{volume}{39}, \bibinfo{number}{8} (\bibinfo{year}{2016}), \bibinfo{pages}{1633--1647}.
\newblock


\bibitem[Yin et~al\mbox{.}(2014)]%
        {yin2014exploring}
\bibfield{author}{\bibinfo{person}{Bin Yin}, \bibinfo{person}{Yujiu Yang}, {and} \bibinfo{person}{Wenhuang Liu}.} \bibinfo{year}{2014}\natexlab{}.
\newblock \showarticletitle{Exploring social activeness and dynamic interest in community-based recommender system}. In \bibinfo{booktitle}{\emph{Proceedings of the 23rd International Conference on World Wide Web}}. \bibinfo{pages}{771--776}.
\newblock


\bibitem[Yu et~al\mbox{.}(2022)]%
        {yu2022surprising}
\bibfield{author}{\bibinfo{person}{Chao Yu}, \bibinfo{person}{Akash Velu}, \bibinfo{person}{Eugene Vinitsky}, \bibinfo{person}{Jiaxuan Gao}, \bibinfo{person}{Yu Wang}, \bibinfo{person}{Alexandre Bayen}, {and} \bibinfo{person}{Yi Wu}.} \bibinfo{year}{2022}\natexlab{}.
\newblock \showarticletitle{The surprising effectiveness of ppo in cooperative multi-agent games}.
\newblock \bibinfo{journal}{\emph{Advances in Neural Information Processing Systems}}  \bibinfo{volume}{35} (\bibinfo{year}{2022}), \bibinfo{pages}{24611--24624}.
\newblock


\bibitem[Yu et~al\mbox{.}(2023)]%
        {yu2023untargeted}
\bibfield{author}{\bibinfo{person}{Yang Yu}, \bibinfo{person}{Qi Liu}, \bibinfo{person}{Likang Wu}, \bibinfo{person}{Runlong Yu}, \bibinfo{person}{Sanshi~Lei Yu}, {and} \bibinfo{person}{Zaixi Zhang}.} \bibinfo{year}{2023}\natexlab{}.
\newblock \showarticletitle{Untargeted attack against federated recommendation systems via poisonous item embeddings and the defense}. In \bibinfo{booktitle}{\emph{Proceedings of the AAAI Conference on Artificial Intelligence}}, Vol.~\bibinfo{volume}{37}. \bibinfo{pages}{4854--4863}.
\newblock


\bibitem[Zhang(2024)]%
        {zhang2024graph}
\bibfield{author}{\bibinfo{person}{Jiahao Zhang}.} \bibinfo{year}{2024}\natexlab{}.
\newblock \showarticletitle{Graph unlearning with efficient partial retraining}. In \bibinfo{booktitle}{\emph{Companion Proceedings of the ACM on Web Conference 2024}}. \bibinfo{pages}{1218--1221}.
\newblock


\bibitem[Zhang et~al\mbox{.}(2021)]%
        {zhang2021multi}
\bibfield{author}{\bibinfo{person}{Kaiqing Zhang}, \bibinfo{person}{Zhuoran Yang}, {and} \bibinfo{person}{Tamer Ba{\c{s}}ar}.} \bibinfo{year}{2021}\natexlab{}.
\newblock \showarticletitle{Multi-agent reinforcement learning: A selective overview of theories and algorithms}.
\newblock \bibinfo{journal}{\emph{Handbook of reinforcement learning and control}} (\bibinfo{year}{2021}), \bibinfo{pages}{321--384}.
\newblock


\bibitem[Zhang et~al\mbox{.}(2019)]%
        {zhang2019comparing}
\bibfield{author}{\bibinfo{person}{Yingxue Zhang}, \bibinfo{person}{S Khan}, {and} \bibinfo{person}{Mark Coates}.} \bibinfo{year}{2019}\natexlab{}.
\newblock \showarticletitle{Comparing and detecting adversarial attacks for graph deep learning}. In \bibinfo{booktitle}{\emph{Proc. representation learning on graphs and manifolds workshop, Int. Conf. learning representations, New Orleans, LA, USA}}.
\newblock


\bibitem[Zhao et~al\mbox{.}(2014)]%
        {zhao2014leveraging}
\bibfield{author}{\bibinfo{person}{Tong Zhao}, \bibinfo{person}{Julian McAuley}, {and} \bibinfo{person}{Irwin King}.} \bibinfo{year}{2014}\natexlab{}.
\newblock \showarticletitle{Leveraging social connections to improve personalized ranking for collaborative filtering}. In \bibinfo{booktitle}{\emph{Proceedings of the 23rd ACM international conference on conference on information and knowledge management}}. \bibinfo{pages}{261--270}.
\newblock


\bibitem[Zhao et~al\mbox{.}(2018)]%
        {zhao2018deep}
\bibfield{author}{\bibinfo{person}{Xiangyu Zhao}, \bibinfo{person}{Long Xia}, \bibinfo{person}{Liang Zhang}, \bibinfo{person}{Zhuoye Ding}, \bibinfo{person}{Dawei Yin}, {and} \bibinfo{person}{Jiliang Tang}.} \bibinfo{year}{2018}\natexlab{}.
\newblock \showarticletitle{Deep reinforcement learning for page-wise recommendations}. In \bibinfo{booktitle}{\emph{Proceedings of the 12th ACM Conference on Recommender Systems}}. \bibinfo{pages}{95--103}.
\newblock


\end{thebibliography}
